# Ultra-fast Graphene-Plasmonic Hybrid Metasurface Saturable Absorber with Low Saturation Fluence

*Md Zubair Ebne Rafique[1,2], Ali Basiri[1,2], Jing Bai[1,2], Jiawei Zuo[1,2], Yu Yao[1,2]\**

[1]School of Electrical, Computer and Energy Engineering, Arizona State University, Tempe, AZ, USA, 85281

[2]Centre for Photonic Innovation, Arizona State University, Tempe, AZ, USA, 85281

**\*Corresponding author: yuyao@asu.edu**

## Abstract

Exploring novel materials with enhanced optical nonlinearities at low power levels with ultrafast response and small footprints is of great interests for information processing, communication, sensing and quantum systems. Recent progress on nonlinear metamaterials and metasurfaces suggests promising solutions to overcome the limitations of nonlinear materials in nature. Here we present a design concept for highly enhanced saturable absorption effect based on subwavelength-thick ($<1/5\lambda_0$) hybrid graphene-plasmonic metasurface structures in infrared wavelengths. Our theoretical and experimental results demonstrated that, by exciting nonequilibrium carriers inside nanoscale hotspots, one could not only enhance the saturable absorption in graphene, but also reduce the saturation fluence by over 3 orders of magnitude (from ~ 1 mJ/cm$^2$ to ~100 nJ/cm$^2$). Our pump-probe measurement results suggested an ultrashort saturable absorption recovery time ($<60$ fs), which is ultimately determined by the relaxation dynamics of photoexcited carriers in graphene. We also observed pulse narrowing effects in our devices based on the autocorrelation measurement results. Such design concepts can be tailored via structure engineering to operate in broader wavelength ranges up to mid-and far- infrared spectral regions. These ultrafast low-saturation fluence saturable absorber designs can enable low-threshold, compact, self-starting mode-locked lasers, laser pulse shaping, and high-speed optical information processing.

## Introduction

Saturable absorbers (SA) are nonlinear optical components with reduced absorption at high optical intensities. They are widely used for mode-locked lasers, Q-switching devices, pulse shaping, optical signal processing applications, etc. Low saturation fluence and ultrafast recovery time are highly desirable attributes of saturable absorbers as they enable access to the nonlinear behavior with lower power consumption and faster speed. Using SA with low saturation fluence and ultrafast recovery time, one can achieve low-threshold, compact, stable, self-starting high-power mode-locked lasers[1], efficient laser pulse shaping for ultrashort pulses[2,3], and high-speed optical signal processing[4–6].

Various saturable absorbers have been demonstrated in literature utilizing different materials and approaches, such as semiconductor saturable absorber mirror (SESAM)[7–10] and quantum dot (QD) saturable absorbers[11–13]. Traditional SESAM and QD require complicated and costly fabrication techniques for precise bandgap engineering, and they have a narrow working wavelength. Carbon nanotube (CNT)[14–17], transition metal dichalcogenide (TMDC)[18–20], topological insulator (TI)[21,22], perovskites[23], black phosphorous[24,25], etc. are also demonstrated as saturable absorbers. CNTs require chirality control for high yield, and other 2D materials such as TMDC and TI show saturation fluence in the range from ~3mJ/cm$^2$ to ~7mJ/cm$^2$ with a few hundred femtoseconds to picosecond recovery time. Reports on planar SA devices such as semiconductor metasurface[6] and plasmonic metasurfaces[26] require high saturation fluences (100 μJ/cm$^2$ to 0.6 J/cm$^2$) with picosecond response time. Graphene-based saturable absorbers[27–35] have attracted much attention because of graphene's broadband optical properties[36,37], nonlinear absorption[32,38], ultrafast recovery time[39–41], and easy integration with various substrates as well as fiber facets[27,29,31,32]. Yet, due to the weak absorption of graphene (~2.3%), weak light-matter interaction with free space light, and ultrashort carrier lifetime, most of the graphene-based saturable absorbers demonstrated so far have high saturation fluence on the order of ~1 mJ/cm$^2$ [27,29,33]. One common method to reduce the saturation fluence is to incorporate a waveguide structure to increase the light-matter interaction with graphene[4,5,42]. Such a method can reduce the saturation fluence to a few mJ/cm$^2$ to μJ/cm$^2$ with a recovery time of 1 ps to 100 fs, yet it usually suffers from high insertion loss (~20 dB) and require large interaction lengths[4,5,42].

In this paper, we demonstrate an ultra-compact graphene-plasmonic hybrid metasurface saturable absorber (GPMSA) experimentally with ultralow saturation fluence (~100 nJ/cm$^2$) and ultrafast recovery time (<60fs) in the infrared wavelength. The subwavelength-thick GPMSA focuses light in the nanogaps between adjacent Au nanobars and significantly enhances the saturable absorption effects in graphene. More specifically, the highly enhanced field intensity in the nanoscale hot spots strongly boosts photon absorption in the graphene over the nanogap regions, resulting in a large photoexcited carrier population. The high density of photocarriers results in a significant change in graphene optical conductivity around the incident wavelengths. Meanwhile, the highly enhanced light-matter interaction[43,44] in these hot spots magnifies the tuning effects of the GPMSA reflection spectra as a result of the change in graphene optical conductivity. Such two-folded enhancement effects finally result in the decrease of saturation fluence by over three orders of magnitude compared to that of free space coupled graphene SA reported in literature[27,29,30] (~1 mJ/cm$^2$). Furthermore, the proposed scheme does not slow down the ultrafast carrier dynamics in the graphene layer. The number of photoexcited carriers rapidly rises upon the arrival of short pulses and relaxes via intraband carrier-carrier and carrier-optical phonon scattering within a few tens of femtoseconds. To the best of our knowledge, the GPMSA devices presented here exhibit the shortest recovery time and lowest saturation fluence among all free space SAs reported in the literature so far[6,8,14–16,18–22,26,27,29,45] (supplementary section S16). The proposed GPMSA holds the promise to enable high-speed signal processing, pulse shaping, and compact high-power pulsed laser sources in infrared wavelengths.

## Results and Discussion

### Design Concept

The proposed GPMSA device consists of a plasmonic metasurface, a dielectric spacer layer ($Al_2O_3$), a metallic back reflector (Aluminum), and a monolayer graphene over the plasmonic metasurface, as shown in Figure 1a. The plasmonic metasurface is composed of gold nanobars closely coupled with each other. At the resonance wavelength of the plasmonic metasurface, the incident light is mainly focused onto the nanogap regions between the adjacent nanobars and results in nanoscale hot spots. The highly enhanced electric field intensity in these hot spots results in strong interaction between the resonance mode of the plasmonic metasurface and the graphene layer on top. By adding a metallic back reflector and a dielectric spacer layer with a thickness of a few tens of nanometers (~20-30nm), we have achieved close to perfect absorption condition[43] at the resonance wavelength. In this condition, no light is reflected, and all the incoming light is absorbed. This can be explained by the modified Fresnel's equations considering multiple reflections of incident light between the top metasurface and the bottom reflector[43,44,46]. By engineering the metasurface design and spacer layer thickness, one can realize destructive interference between the first partially reflected wave and the following higher-order reflected partial waves at the metasurface to achieve close to 100% absorption in the metasurface cavity (see supplementary information S1). Such a metasurface absorber design configuration can further enhance the light graphene interaction and results in up to 3 orders of magnitude enhancement for the near field intensity in the nanoscale hotspots, as shown in the inset of Figure 1a. Full-wave simulation results showed that the optical absorption in the graphene layer over the nanogap region (100 nm ×30 nm) at resonance wavelength (~1.035μm) is close to 40% of all the incident light inside the device unit cell (150 nm × 450 nm), as shown in Figure 1b (more simulation details are provided in supplementary information S6), in comparison with ~2.3% in a suspended monolayer graphene[47].

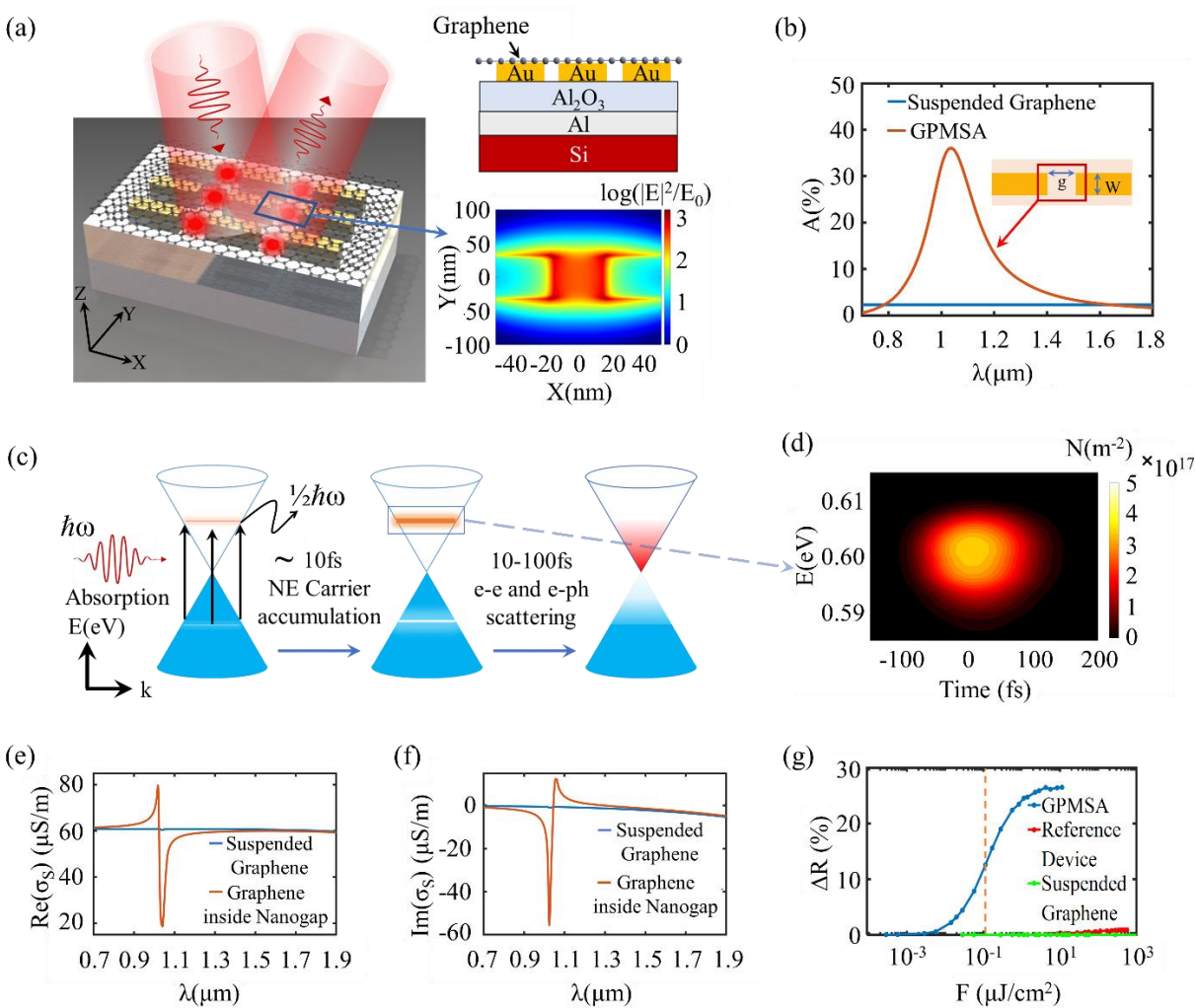

***Figure 1. Graphene-plasmonic hybrid metasurface saturable absorber (GPMSA) design concept***

*(a) A schematic of the GPMSA device (left), its cross-sectional view (top right), and the near field intensity enhancement near the nanogap between adjacent Au-nanobars at resonance wavelength*

*(1.035μm) of the GPMSA (bottom right). $|E|^2$ is the electric field intensity induced between the adjacent nanobars, and $E_0^2$ is the intensity of the incident electric field. GPMSA device parameters used in all simulations presented in this manuscript: Au nanobar length, L=120nm, width, W=100nm, gap, g=30nm, thickness, $t_{Au}$=40nm, x-period, $P_x$=150nm, y-period, $P_y$=450nm and $Al_2O_3$ thickness, $t_{AlO}$=40nm. (b) The simulated overall Absorption, A (%), of the incident laser light in graphene inside the nanogap region (100nm x 30nm) of GPMSA compared to that of a suspended graphene layer without the metasurface structure. Both simulations were performed over the same area as the unit cell of the plasmonic metasurface ($P_x$=150 nm, $P_y$=450 nm) (c) Carrier dynamics in graphene under ultrafast laser excitation with photon energy $E_{ph}$= ħω. Photoexcited electrons and holes generated by the incident photons form a non-equilibrium (NE) carrier distribution centered at energy ħω/2 in the conduction band (CB) and valence band (VB). Within 10 – 100 fs, the non-equilibrium carriers relax to a hot fermi-Dirac distribution due to electron-electron (e-e) and electron-optical phonon (e-ph) scattering. (d) Simulated 2D contour plot of the photoexcited non-equilibrium electron distribution, $N(m^{-2})$, in the graphene over the GPMSA nanogaps at energies around $\frac{1}{2}$ħω, i.e., from 0.585eV to 0.615eV above the conduction band edge ($E_C$=0eV). The pump fluence is 112nJ/cm². For this and the following simulations in the manuscript, we use the same laser parameters as our measurement setup: pulse width 110fs; repetition rate 100MHz; central wavelength 1.035μm, and spectral broadening ~17nm. The x-axis represents the time delay with respect to the peak of the incident light pulse. (e) The instantaneous change in the real part Re(σ_S) and (f) the imaginary part of graphene optical surface conductivity Im(σ_S) of the GPMSA device presented in Figure 1d, at the peak of the incident pulse (delay time=0 in Figure 1d), in comparison with that of a suspended graphene layer without metasurface structures. (g) Simulated reflection modulation ΔR (%) of the same device in Figure 1a at the peak*

*intensity of different incident pump fluences for GMPSA, the reference device, and suspended graphene.*

When an ultrashort laser pulse incident onto the GPMSA device, the plasmonic metasurface focuses most of the optical energy into the "hot spots" in the nanogaps. Close to 40% of the incident photons are absorbed in the graphene located at the hotspots (as shown in Figure 1b) and generate a large population of transient photoexcited non-equilibrium (NE) carriers within a few tens of femtoseconds. The process is illustrated in Figure 1c. The electrons in the valence band absorb the incident photons, get excited to the conduction band, and form a non-equilibrium carrier distribution around the energy band covered by the laser wavelength and its spectral bandwidth. These photoexcited non-equilibrium carriers relax into a hot fermi-Dirac distribution within 10 – 100 fs due to electron-electron (e-e) and electron-optical phonon (e-ph) scattering[30,41]. We adopted a model considering this non-equilibrium photocarrier accumulation upon laser excitation and their subsequent relaxation due to electron-electron and electron-optical phonon scattering[30]. The dynamic interplay between the occupation probability of electrons in the conduction band ($f_C(t, \omega)$) and the valence band ($f_V(t, \omega)$) upon laser excitation is explained by two semi-empirical coupled differential equations relating the density of states of graphene at the excitation photon energy (D(E)), incident light intensity (I (t, $\omega$)) and the associated relaxation time, $\tau_1$, which are provided below[30],

$$\frac{\partial f_V(t,\omega)}{\partial t} = - f_V(t,\omega)\ \frac{\pi \alpha_f}{D(E)\ \hbar\omega}\ I_0 exp(-\frac{t^2}{\Delta^2})\ + f_C(t,\omega)\ \frac{\pi \alpha_f}{D(E)\ \hbar\omega}\ I_0 exp(-\frac{t^2}{\Delta^2}) + \frac{1 - f_V(t,\omega)}{\tau_1} \tag{1}$$

$$\frac{\partial f_C(t,\omega)}{\partial t} = f_V(t,\omega)\ \frac{\pi \alpha_f}{D(E)\ \hbar\omega}\ I_0 exp(-\frac{t^2}{\Delta^2}) - f_C(t,\omega)\ \frac{\pi \alpha_f}{D(E)\ \hbar\omega}\ I_0 exp(-\frac{t^2}{\Delta^2}) - \frac{f_C(t,\omega)}{\tau_1} \tag{2}$$

Where, $I(t) = I_0 exp(-2.76t^2/\Delta^2)$ is the incident laser intensity, $\alpha_f$ is the fine structure coefficient of graphene, $D(E) = D(\hbar\omega/2) = \hbar\omega/(\pi\hbar^2 v_F^2)$ is the density of states of graphene on the CB or VB at incident photon energy, and $\Delta$ is the incident laser pulse width (Full Width Half Maxima-FWHM). These two coupled differential equations explain the absorption, stimulated emission, and the subsequent relaxation of the excited non-equilibrium carriers. For saturable absorption, we are primarily interested in the ultrafast accumulation of non-equilibrium photoexcited carriers from VB to CB and the subsequent relaxation of the carriers from the non-equilibrium states to the equilibrium Fermi-Dirac distribution near the Dirac point. The most contributing relaxation channels in this process are the carrier-carrier (e-e) and carrier-optical phonon (e-ph) scattering, which happens within 10-100 fs. The scattering contributions from other origins such as carrier-acoustic phonon and phonon-phonon scattering are not considered here as they have a picosecond or larger time constant[40]. So, the relaxation time, $\tau_1$, only consists of the excited carriers' intraband electron-electron and electron-optical phonon scattering rates. The detailed simulation process of the model can be found in supplementary information S2. Based on this model, we performed time-dependent FDTD simulations of the GPMSA device, in a self-consistent way, upon ultrashort laser pulse excitation based on an iterative method, taking into account the dependence of optical absorption in the GPMSA device on the incident light intensity (supplementary information S3). For each incident pulse fluence, we obtain the time-dependent photocarrier distribution and graphene surface conductivity from the model and the corresponding reflection spectra of the GPMSA is obtained from FDTD simulations. Figure 1d shows the obtained time-dependent photoexcited electron populations, $N(m^{-2})$, around half of incident photon energy (from 0.585eV to 0.615eV) above the conduction band edge ($E_C$=0eV) in the graphene over the nanogap regions of the GPMSA device, in Figure 1b, for an incident laser pulse. The peak photocarrier populations generated in the nanoscale hot spots are two orders of magnitude larger than that in the suspended graphene layer (Figure S11). As a result of the transient photocarriers, the optical surface conductivity of graphene is modulated with respect to time and wavelength. Figure 1e and 1f show that both the real part and imaginary parts of the optical surface conductivity for the graphene located in the nanoscale hotspots are strongly modulated ($\Delta\sigma_S$ = -(42 - 57i) μS/m) at resonance wavelength compared to that of a suspended graphene layer under the same pulse fluence ($\Delta\sigma_S \approx$ -(0.8 - 1i) μS/m). Besides the significantly increased modulation of graphene optical conductivity, the reflection of the GPMSA device is also highly sensitive to the change of graphene optical surface conductivity inside the nanoscale hotspots[43]. We performed full-wave simulations (in a self-consistent way) to evaluate its reflection modulation, $\Delta R$, at different pump fluence. Note that the reflection modulation $\Delta R$ is defined as the difference between the maximum reflection upon incident pulse ($R_{max}$) and the reflection with no pump ($R_0$), i.e., $\Delta R$ (%) = ($R_{max}$ - $R_0$) x100%. Figure 1g provided the corresponding reflection modulation of the GPMSA mirror at different pump fluence compared to a reference device composed of monolayer graphene on the same substrate ($Al_2O_3$/Al/Si) as the GPMSA device but without plasmonic metasurfaces and the suspended graphene layer. The reference device acts as a saturable mirror with a metallic back reflector, similar to the GPMSA devices, and thus is more appropriate for evaluating performance enhancement by the metasurface design than the suspended graphene layer. According to the simulation results in Figure 1g, the reflection of the GPMSA mirror is increased by ~15% at the peak of the incident pulse (indicated by the dotted line) for the pump fluence of $112nJ/cm^2$ while the reflection spectra of a reference device remain unchanged for the same pump fluence since it is much lower than the saturation fluence of the reference device. By fitting the simulated reflection modulation with a two-level atom model[48], the saturation fluence of the GPMSA mirror is found to be ~120nJ/cm² which is close to three orders of magnitude smaller than that of the reference device (~100μJ/cm²) and more than three orders of magnitude smaller than that of the suspended graphene layer (~1mJ/cm²), also shown in Figure 1g. Therefore, the GPMSA device acts as a saturable mirror at significantly reduced pumping fluence compared to the reference device. Furthermore, the maximum reflection modulation of the GPMSA mirror is ~30%, more than one order of magnitude higher than that of the reference device.

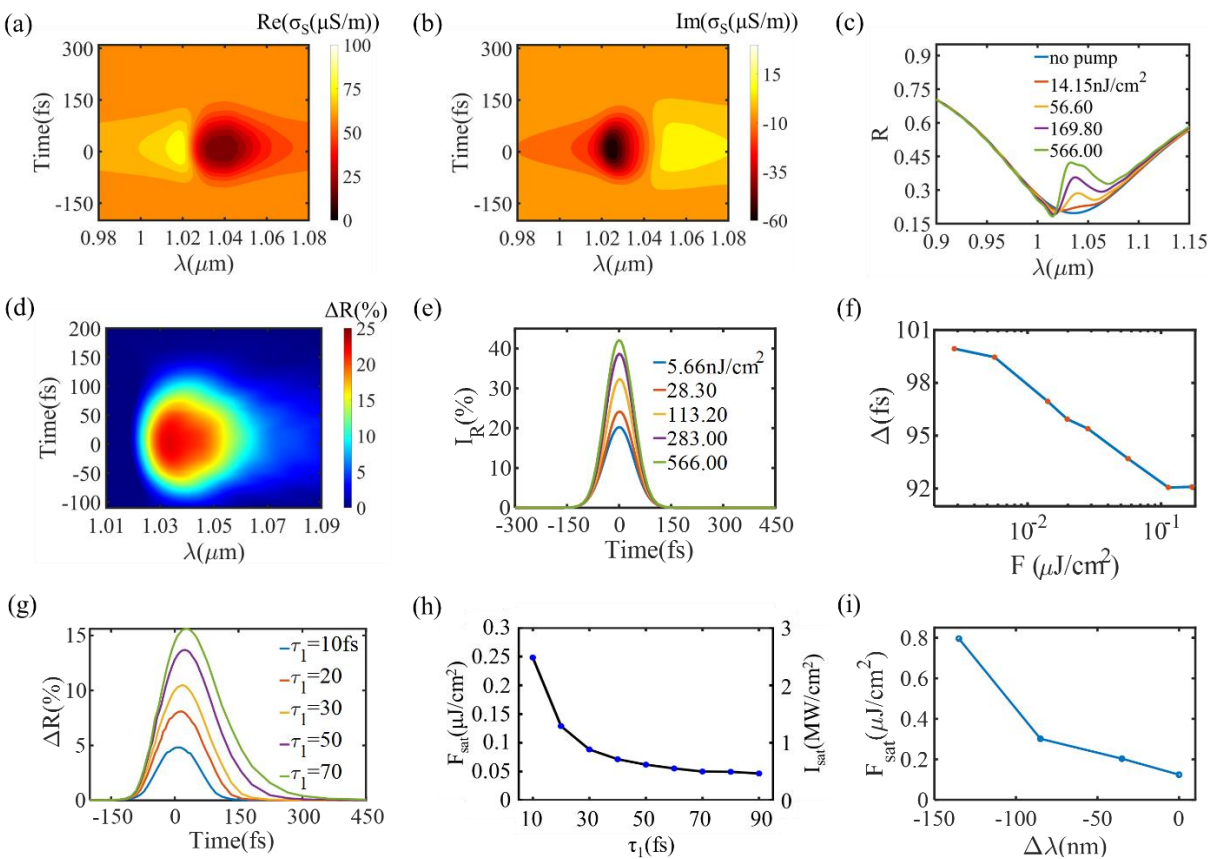

*Figure 2: **Simulation results of GPMSA devices.** (a, b) Simulated modulation of optical surface conductivity, $\sigma_S$ ($\mu S/m$), (a-real and b-imaginary part) of graphene inside the nanogap of GPMSA (structure parameters for this simulation and all simulations in the manuscript are the same as presented in Figure 1a) with respect to time and wavelength after pump excitation with fluence 112nJ/cm². The laser parameters used in the simulations presented here are the same as in Figure*

1d. (c) Simulated reflection spectra of GPMSA for different incident pump fluences (at the peak of the pulses) at laser wavelength 1.035μm compared to no pump condition. Note that the GPMSA device used in all the simulations of this manuscript has a resonant wavelength of 1.035μm. (d) Simulated reflection modulation, ΔR (%), of GPMSA with respect to wavelength and time at 566nJ/cm² incident pump fluence. ΔR is ~25% at the laser pulse peak intensity around the laser center wavelength (1.035μm). (e) The simulated amplitude of the reflected pulse, $I_R$ (%), with respect to time for increasing incident pump fluences. (f) Simulated pulse width, Δ (fs), of the reflected pump pulses at the laser center wavelength for different incident pump fluence. (g) Simulated reflection modulation, ΔR (%), for different carrier relaxation times, $τ_1$, in graphene. (h) Simulated saturation fluence, $F_{sat}$, and peak saturation intensity, $I_{sat}$ (MW/cm²), of GPMSA at different graphene carrier-relaxation times, $τ_1$. (i) Saturation fluence of GPMSA device as a function of wavelength difference between the incident light and the device resonance, Δλ.

Figure 2a and b show the simulated change in the real and imaginary part of the graphene's optical surface conductivity ($σ_S$), with respect to time and wavelength, inside the nanogaps of GPMSA (structure parameters are the same as in Figure 1a, resonant at 1.035μm wavelength) for laser excitation with pump fluence 112nJ/cm² at 1.035μm laser wavelength. As a result of the changes in graphene's optical surface conductivity, the reflection spectra of GPMSA change. Figure 2c shows the simulated reflection spectra of the device at the peak of the laser pulses (time = 0fs) for different incident pumping fluences. The reflection spectra change of GPMSA is also time-dependent as graphene's optical surface conductivity changes with time. This is shown clearly in Figure 2d, which illustrates the time and wavelength-dependent reflection modulation, ΔR (%), of GPMSA at 566nJ/cm² pumping fluence (a similar figure for the reference device showing <1% reflection modulation for even larger pump fluence around 5.66μJ/cm² is shown in Figure S12).

From Figure 2c, we see that the reflection modulation is dependent on the incident pump fluence. So, to evaluate the device's operation as a saturable absorber, the self-modulation of reflection, $\Delta R(\%)$, at the center wavelength of the incident laser and at the peak of the pulse (time = 0fs) for different incident pump fluences are plotted in Figure 1g (blue curve), showing the saturation effect of the device.

Theoretical analysis of the reflected pulse widths from the GPMSA devices suggested pulse narrowing effects. Figure 2e presents the simulated reflected pulse amplitudes at different incident pumping fluences. It is obtained considering the device's self-modulation effect (Figure 2d), and a normalized Gaussian incident pulse, $I_{in} = I_0 \, exp(-\frac{t^2}{\Delta^2})$ with pulse width $\Delta$ =110fs and amplitude $I_0$=1. The corresponding pulse widths, $\Delta$(fs), of the reflected pulses at different pump fluence are shown in Figure 2f. The reflected pulse width decreases with increasing pump fluences. The significant decrease (10%) of reflected pulse width relies on two important features of the GPMSA, i.e., saturated absorption at high intensity and ultrafast recovery time (shorter than half of the pulse width ~55fs), because both are necessary to ensure the peak of the pulse with highest intensity experiences larger reflection than the rising and falling edges of the pulse. If the recovery time is longer than the pulse width, the absorption experienced by the falling edge of the pulse would be close to that at the peak intensity, thus, the pulse width narrowing effect will be much weaker.

An essential factor for the GPMSA operation is the carrier relaxation time of graphene, i.e., $\tau_1$. This is defined as the time necessary for the photoexcited non-equilibrium carriers to relax from the excited states with the help of e-e and e-ph scattering events (other longer time scale relaxation channels are not considered here, as explained in the previous section and in supplementary information S2 and S3). Reflection modulation, saturation fluence, and recovery time of GPMSA are dependent on $\tau_1$. Figure 2g shows that for the same incident pump fluence (56.6nJ/cm$^2$), shorter or longer $\tau_1$ will result in small or large reflection modulation, respectively. This is understandable because, for slower $\tau_1$, the non-equilibrium carriers take longer to relax, i.e., they stay longer at the excited states, and it helps to build up the number of carriers in non-equilibrium states. As the non-equilibrium carrier concentration increases, graphene optical surface conductivity goes through larger changes, resulting in larger modulation of the device response. For faster $\tau_1$, carriers are relaxed from non-equilibrium states quickly and have less time to build up. So, the change in graphene optical surface conductivity and the consequent change in device reflection spectra is smaller for faster $\tau_1$ compared to slower $\tau_1$. As a result, for faster $\tau_1$, larger pump fluence is required to obtain similar modulation as the slower $\tau_1$. Consequently, obtaining large reflection modulation with an ultrafast device is very challenging.

The saturation fluence, $F_{sat}$, of GPMSA also varies for different relaxation times, $\tau_1$, illustrated in Figure 2h. For faster $\tau_1$, to saturate the device, we need higher pump fluence, whereas, for slower $\tau_1$, the required pump fluence is smaller. This observation explains the fundamental reason behind the difficulties in reducing saturation fluence in graphene while maintaining ultrafast carrier relaxation time ($F_{sat}$ vs. $\tau_1$ for reference device is shown in figure S13). Furthermore, we also studied the case for pump laser wavelengths different from the device's resonance wavelength (1.035μm). The saturation fluence increases quickly as the incident pump laser wavelength shifts far away from the device resonance, as shown in Figure 2i. This is due to lower near-field enhancement inside the nanogaps of the Au nanobars for off-resonance pumping wavelengths (Figure S15a). This, in turn, leads to decreased photon absorption in graphene, and thus produces smaller modulation in its optical conductivity. Therefore, the saturation fluence of the device becomes higher (Figure 2i) and reflection modulation becomes lower (Figure S15b, c) as the pumping wavelength is away from the resonance, resulting in a limited operation spectral range of the saturable absorber. We will continue exploring designs with a broader operation wavelength range.

**Fabrication and characterization of GPMSA devices**

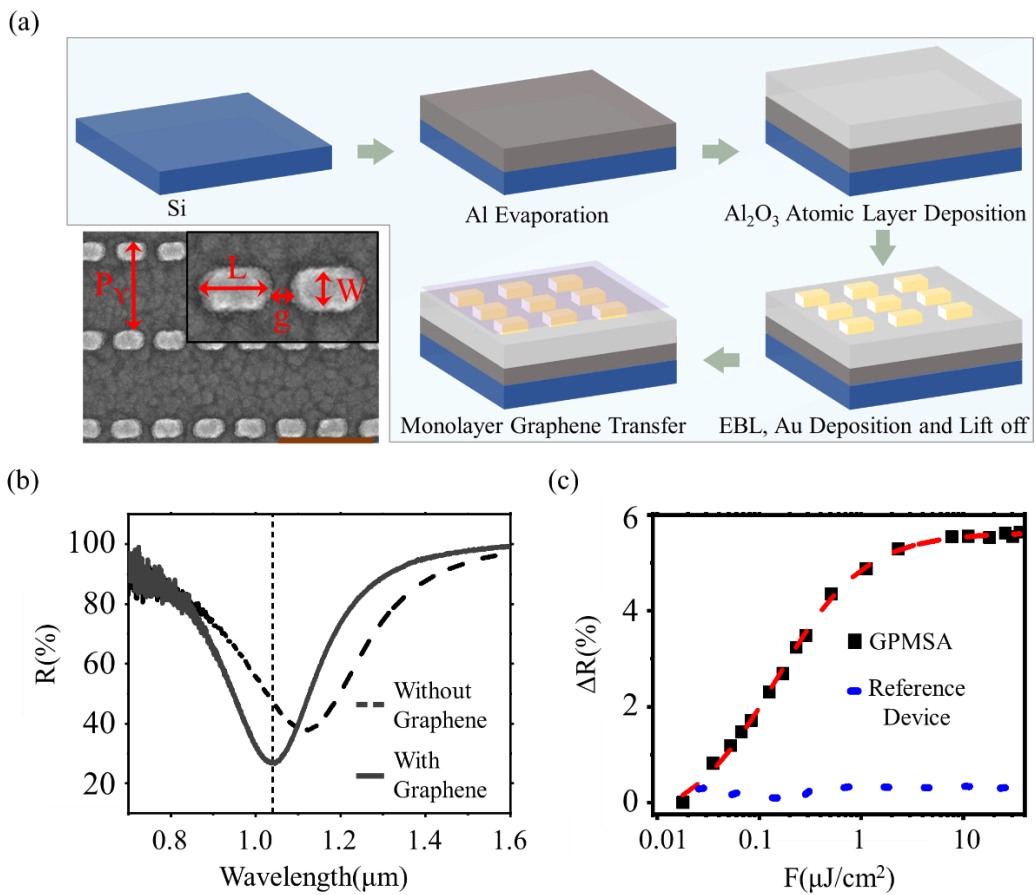

**Figure 3. Fabrication and characterization of GPMSA devices.** *(a) Device fabrication process and a scanning electron microscope (SEM) image of the device after fabrication (before graphene transfer). The scale bar (500nm) is indicated with a solid brown line on the lower right corner. The inset shows a close-up view of two nanobar structures. Device Parameters are L=150nm, W=100nm, g=46nm, $P_Y$=470nm, $t_{Au}$=40nm, and $t_{AlO}$=20nm. Here, L is the length of the Au nano bar, W is the width of the nano bar, g is the gap between the coupled nano bars in the x-direction,*

and $P_y$ is the period of the structure in the Y direction, $t_{Au}$ is Au nano bar thickness and $t_{AlO}$ is $Al_2O_3$ thickness. (b) Measured reflection spectra of the device (presented in Figure 3a) before and after graphene transfer. The resonance wavelength of the device is at 1.035μm. (c) Measured reflection modulation, ΔR (%) at 1.035μm, of the GPMSA device compared to the reference device at different pump fluences. Red dash lines are the fitted curves.

We fabricated the device on a Si substrate. First, 150nm Aluminum was evaporated (using electron-beam evaporator) as the back reflector. Then 20nm $Al_2O_3$ dielectric spacer layer was deposited on top of Al using Atomic Layer Deposition (ALD). The Au plasmonic metasurface on the dielectric layer was fabricated using a double layer resist of poly (methyl methacrylate) (495K PMMA with 2% Anisole, 100nm) and methyl methacrylate (950K MMA with 2% Anisole, 70nm). Electron Beam Lithography (EBL) was used to pattern the double layer resist. After patterning the sample, cold development (4ºC temperature) in a solution with 1-part Methyl Isobutyl Ketone (MIBK) and 3-part Isopropyl alcohol (IPA) was done. Then Cr/Au (5nm/35nm) layer was deposited using a thermal evaporator, and lift-off was done using hot acetone for 12 hours. PMMA/MMA bilayer resist was used to form the overhang resist structure after development and facilitate the lift-off of the plasmonic metasurface structure. The fabrication flow is shown in Figure 3a. An SEM image of the device, taken before graphene transfer, is shown in Figure 3a. Monolayer graphene was transferred on the device using a wet transfer method (see the methods section). We performed Raman spectroscopy to quantify graphene layer and its quality. The measured spectra are shown in Figure S16. Raman spectra reveal that the 2D/G peak is larger than two, indicating monolayer graphene, and a very small D peak indicates good quality. We used a FTIR spectrometer coupled with a microscope, linear polarizer, and a liquid nitrogen-cooled Mercury Cadmium Telluride (MCT) detector (Figure S17a) to measure the reflection spectra of the GPMSA devices. Figure 3b shows the reflection spectra before and after the graphene transfer for a GPMSA device (SEM image and device structure parameters are presented in Figure 3a). After graphene transfer, the resonance wavelength of the device was blue-shifted by ~76nm, the reflection dip is decreased by 11%, and the reflection at the working wavelength (~1.035μm) is changed by 20%. Furthermore, we can tailor the working wavelengths of the GPMSA design (Figure S17b) by changing the geometric of the devices, i.e., length, width, gap, and spacer layer thickness.

The optical setup for the saturable absorption measurement (shown in Figure S18a) comprises a femtosecond laser (Menlo System GmbH) with 110fs pulse width and 100MHz repetition rate at center wavelength 1.035μm and spectral broadening of ~17nm. The laser's output power can be changed from 0 to 1200mW (~65μJ/cm$^2$) by varying pump diode currents. The diameter of the focus point on the sample was around 150μm (supplementary information S13). The fabricated GPMSA device area was 200μm x 200μm, so the laser beam was focused on the center of the device to capture most of the incoming light intensity. After the incident light was reflected off the sample, it was captured by a power meter. We recorded the reflected power values from GPMSA and from the aluminum reflector without the graphene-plasmonic metasurface. These two values were used to calculate the reflection of the samples (GPMSA and reference device) at 1.035μm. The measured reflection of GPMSA (same device as in Figure 3b) at 1.035μm increased for increasing pump fluence and reaches a saturation level after it goes beyond the saturation fluence of the device (Figure S18b). This behavior is a characteristic indicator of saturable absorption property. Net reflection modulation, $\Delta R$ (%), of this device is more than 5.5%, and the relative modulation depth ($\Delta R/R_0$) is around 20% at 5μJ/cm$^2$ incident pump-fluence. Consequently, the insertion loss of GPMSA is ~4.25dB. Compared to the state-of-the-art of graphene-based SA[4], the insertion loss is four times smaller.

Figure 3c shows the measured net reflection modulation, ΔR (%), profile (ΔR vs. incident pump fluence) for the GPMSA device (same as in Figure 3b) compared to the reference device. The GPMSA device exhibits much higher reflection modulation, >5.5%, compared to almost zero (i.e., unchanged) reflection modulation of the reference device. The measured reflection modulation profiles of similar GPMSA devices (resonance at 1.035μm) show quite consistent saturable absorption properties (Figure S19). Almost zero reflection modulation from the reference device indicates that there is no nonlinear absorption in the $A_2O_3$/Al/Si stack. Note that the femtosecond laser's spectra blue-shift due to output power change (Supplementary section S17 and Figure S22) does not affect our measurement results as it introduces negligible reflection modulation (~0.1%) unrelated to saturable absorption compared to the large reflection modulation (~5.5%) obtained from the saturable absorption measurement of the device.

To extract the device's saturation fluence, we followed the two-level atom model[48] and used the following equation to fit the reflection profile (vs. incident pump fluence) of GPMSA.

$$R = 1 - (A_{NS} + \frac{A_s}{1 + \frac{F}{F_s}}) \tag{3}$$

Here, $A_{NS}$ describes the non-saturable absorption, $A_s$ is the saturable absorption. $F_s$ is the saturation fluence within the single photon absorption limit. After fitting the measured data with equation (3), we obtained the saturation fluence ($F_s$) of the device. The fitted curve is shown in Figure 3c with red dashed lines. The estimated saturation fluence from this fitting is ~100nJ/cm$^2$. The saturation fluence is 2-3 orders of magnitude smaller than the graphene SA reported in literature[27,29].

**Ultrafast saturable absorption in GPMSA devices**

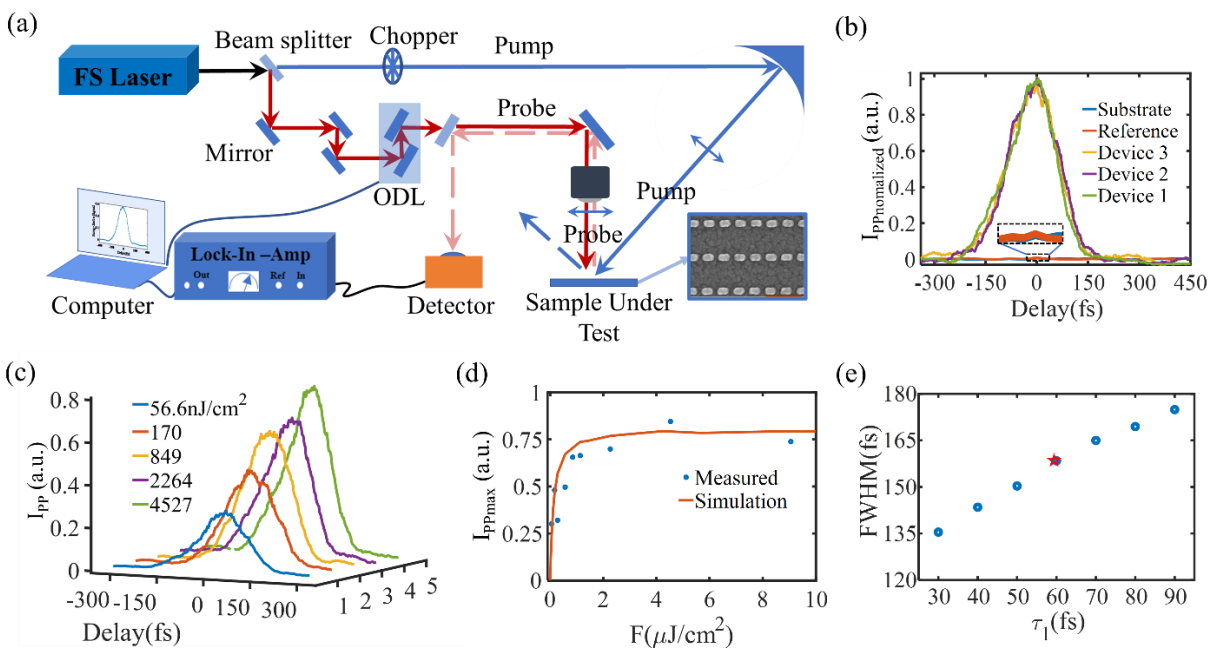

***Figure 4. Degenerate Pump-Probe Measurement of GPMSA*** *(a) Pump-probe measurement setup comprising of femtosecond laser working ~1.035μm, beam splitter, optical delay line, focusing optics and photodetector. A chopper and lock-in amplifier are used to increase the measurement's signal-to-noise ratio (SNR). (b) Measured pump-probe signal (normalized, $I_{PPnomalized}$) from the substrate (Al₂O₃/Al/Si), reference device, and three GPMSA devices (same devices as presented in Figure 3c (Device 1) and Figure S19a) at 566nJ/cm² pump fluence. (c) Measured pump-probe signals at different pump fluences for GPMSA (Device 1). (d) Measured and simulated pump-probe signal peaks with respect to incident pump fluences (Device 1). (e) Full width half maximum (FWHM) of the simulated pump-probe signals (blue circles) at different carrier relaxation times $\tau_1$ and the measured pump-probe signal (red star) at 56.6nJ/cm² pump fluence.*

To investigate the transient behavior and obtain the recovery time of GPMSA, we have performed degenerate pump-probe measurements at 1.035μm wavelength with the femtosecond laser (Menlo System GmbH) as mentioned previously. A schematic of the optical setup is shown in Figure 4a (details can be found in the methods section). With the optical delay line (ODL), we varied the delay between the pump and probe pulses and recorded the change in the device reflection with the probe pulse as the device interacted with the pump pulse. Pump-probe measurements show that GPMSA devices provide enhanced absorption and large reflection modulation within femtosecond time scale compared to the substrate and reference device (Figure 4b). The maximum reflection modulation occurred around zero-time delay between the pump and probe pulses. It is noticeable that the rising part is longer than the decaying part of the pump-probe signal. We think it is likely due to the fact that on the rising edge, two competing processes are happening at the same time within the excitation energy band (i.e. around $\frac{1}{2}\hbar\omega \approx 0.6\text{eV}$) of graphene conduction band: absorption of photons (photocarrier generation) and e-e scattering of the excited photocarriers (photocarrier relaxation), while the falling edge is dominated mainly by the photocarrier relaxation. Because of these two competing processes on the rising edge, there is a slower increase of nonequilibrium carrier concentration and thus takes longer time to reach the maximum reflection modulation. This trend was also observed in the simulation presented in Figure 2g and Figure S14. The ultrafast reflection modulation also depends on the incident pump fluence, as shown in Figure 4c. The reflection modulation goes up as the pump fluence is increased. The maximum values of the measured pump-probe signal at different pump fluences, Figure 4d, reveal saturation of the signal at higher pump fluences. The measurement results showed reasonably good agreement with simulation results (Figure 4d) obtained based on the model presented in the previous section.

To extract the recovery time of the device, we first fitted the decaying portion of the pump-probe signal with an exponential decay function to obtain the recovery time, but the exponential decay function does not fit well with the measured pump-probe signal. This happens because the recovery time of the device is dominated by the ultrafast carrier relaxation (for the photoexcited non-equilibrium carriers) of graphene, which happens within 10-100fs (<100fs) due to e-e and e-ph scattering events, which is shorter than the pulse widths of the pump and probe lasers. Thus the pump-probe signal cannot provide the actual relaxation time of the GPMSA device and produce a non-exponential decay of the signal. For this reason, we performed time-dependent FDTD simulations of GPMSA, in a self-consistent way, utilizing the time-dependent optical surface conductivity of graphene for different carrier relaxation times, $\tau_1$, obtained from the model described in the previous section (Supplementary Information S3). Figure 4e shows the comparison between the FWHM of the measured pump-probe signal (~160fs at 56.6nJ/cm$^2$ pump fluence) and that of the simulated pump-probe signal for different carrier relaxation times. The results suggest that the carrier relaxation time is less than 60fs. By comparing the simulation results with the measured pump-probe signal, we found that the time-dependent simulations with 50fs to 60fs recovery times show good agreement with the measured pump-probe signal (Figure S20). At such an ultrashort time scale, the relaxation time of excited non-equilibrium carriers in graphene dominates the recovery time of our device, so we can conclude that our device has a recovery time of less than <60fs. Our device shows a faster recovery time than most reported works, including all graphene-based saturable absorbers, in the literature (Figure S21). We attribute it to the ultrafast nonequilibrium carrier relaxation time in graphene. Moreover, it is likely that the plasmonic metasurface provides new relaxation channels around the highly enhanced near-field hot spots in the nanogaps between adjacent antennas, which helps the hot carrier population and optical phonon population in graphene to relax faster. Further investigations are necessary to gain better insights into near-filed assisted carrier relaxation processes.

**Autocorrelation: pulse-width measurement**

Pulse narrowing is one of the important functions of saturable absorbers[49]. According to the simulation results, we expect decreased pulse width as incident fluence increases (Figure 2f). In experiment, we performed ultrashort pulse autocorrelation measurements to investigate the pulse width narrowing effect of the GPMSA devices. The measurement setup is shown in Figure 5a. Fs-laser pulses were incident on the GPMSA device and the reflected light was coupled into a two-photon-based interferometric intensity autocorrelator for femtosecond lasers (FSAC) to measure the pulse width. A simple illustration of the pulse width narrowing concept is presented in Figure 5b, where we expect to see a narrowing effect of the reflected pulse width from the GPMSA device (incident pulse width, $\Delta_{in}$, and reflected pulse width, $\Delta_{Ref}$). Figure 5c shows the intensity correlation as a function of the time delay between the pulses from the delay arm and reference arm inside the autocorrelator, obtained for the reflected pulse by the GPMSA, the reference device as well as the substrate back reflector (without graphene) at the same pump fluence ~58μJ/cm². The autocorrelation curve for the GPMSA device is narrower compared to those for the reference device and the substrate back reflector. This, together with experimental results in Figures 3 and 4, confirms the ultrafast nonlinear absorption and the resulted pulse narrowing of the GPMSA devices, which did not happen in either the reference device or the substrate back reflector. Figure 5d shows the measured reflected pulse widths from the GPMSA device and the incident pulse widths at different pump fluences. Note that the output pulse width from the fiber laser source decreases as the laser pulse energy increases. Nonetheless, at all incident fluence levels above the saturation fluence, the pulse width becomes narrower after reflection from the GPMSA than the input pulse width by about 10%, which agrees with the theoretical analysis presented in Figure 2f.

The results again confirm that the GPMSA device has strong saturable absorption with an ultrafast recovery time.

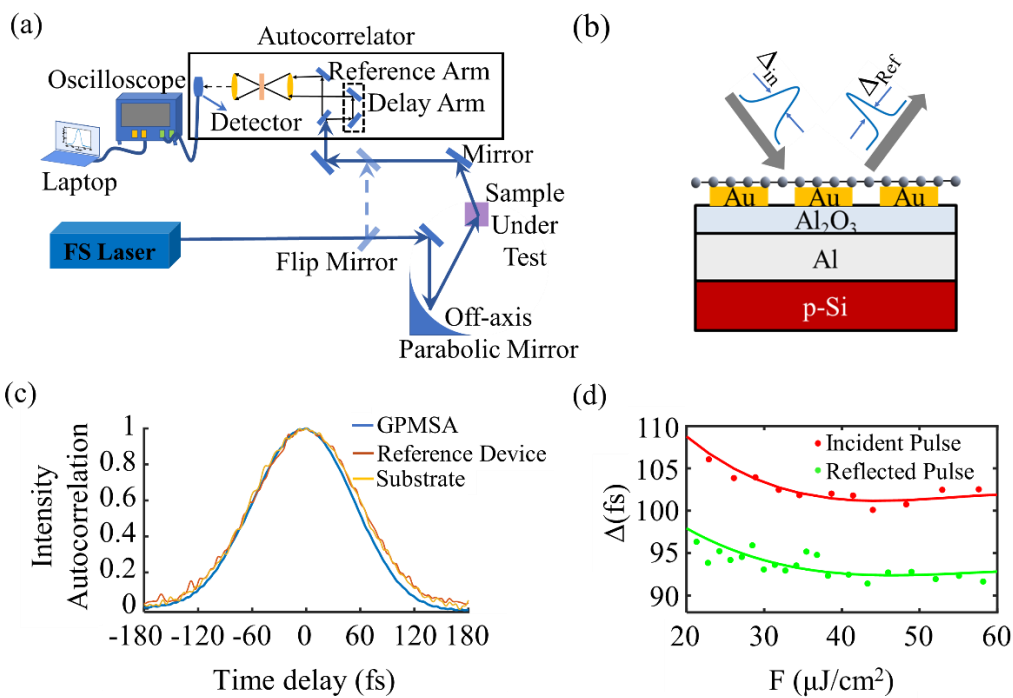

***Figure 5. Reflected pulse width measurement.*** *(a) Pulse width measurement setup comprising of the femtosecond laser, focusing optics and autocorrelator. The intensity autocorrelator is used to measure the pulse width. (b) Schematic representation of the pulse narrowing effect after the laser pulse is reflected from the GPMSA device. (c) The pulse shapes, intensity autocorrelation, of the reflected laser light from GPMSA (Device 1 in Figure 4b) compared to the reference device and substrate at pump fluence ~58 μJ/cm². (d) Comparison of the reflected laser pulse width, $\Delta_{Ref}$, from GPMSA (Device 1 in Figure 4b) and incident pulse width, $\Delta_{in}$, at different incident laser pulse fluence. For input pulse fluence smaller than 20 μJ/cm², the autocorrelation measurement results could not provide sufficient signal-to-noise ratio for pulse width estimation.*

## Conclusion

In summary, we have demonstrated ultra-fast saturable absorption in subwavelength-thick (~200 nm, $< \lambda_0/5$) GPMSA devices based on graphene-plasmonic hybrid metasurfaces. Both theoretical and experimental results demonstrated that the GPMSA devices are featured with ultralow saturation fluence (~100 nJ/cm$^2$) and ultrafast recovery time (<60fs) in the infrared wavelengths. The GPMSA devices focus light into the nanoscale hot spots inside the nanogap between adjacent Au nanoantenna and significantly increase light-matter interaction with graphene. Consequently, photon absorption in the graphene is significantly boosted over the nanogap regions and generates a large photoexcited carrier population. Furthermore, the strong light-graphene interaction in the hot spots of the GPMSA magnifies the tuning effects of the reflection spectra due to the ultrafast modulation of graphene optical conductivity. These enhanced effects finally result in the decrease of saturation fluence by over three orders of magnitude compared to free space coupled graphene SAs reported in the literature. Moreover, the proposed scheme does not slow down the ultrafast carrier dynamics in the graphene layer while the saturation fluence of the device is decreased. To the best of our knowledge, the GPMSA devices presented here exhibit the shortest recovery time and lowest saturation fluence among all free space SAs reported in the literature so far. The design concept can be readily applied to longer infrared wavelengths via structure engineering thanks to the broadband absorption of graphene and the design flexibility of plasmonic metasurfaces. The proposed planer hybrid metasurface-based GPMSA design holds the promise to enable high-speed signal processing, pulse shaping, and compact high-power pulsed laser sources in the infrared wavelengths with low power consumption.

## Methods

### Numerical Simulation

All numerical simulation is done using MATLAB® (MathWorks Inc.) with a phenomenological graphene carrier dynamics model[30] upon laser excitation, considering the non-equilibrium carrier accumulation and subsequent relaxation of the non-equilibrium carriers. The detailed simulation process can be found in the supplementary information S2 and S3.

### FDTD Simulation

All FDTD simulations are done using commercially available Lumerical Inc. FDTD software. The optical surface conductivity of non-pumped graphene is extracted from the numerical model included in the software, and the time-dependent optical surface conductivity of pumped graphene is calculated using the numerical model described above. All other material parameters (Gold, Aluminum, and Aluminum Oxide-$Al_2O_3$) are also obtained from the simulation software database. Time-dependent FDTD simulation process details can be found in the supplementary information.

### Plasmonic metasurface fabrication

On a Silicon substrate, a 150nm Aluminum (Al) layer was deposited using electron beam evaporation (PVD 75, Kurt J. Lesker Company®). On top of the Aluminum layer, a 20nm Aluminum Oxide ($Al_2O_3$) layer is deposited using atomic layer deposition (Cambridge Savannah ALD) as a dielectric spacer layer. The dielectric layer was spin-coated with two layers of electron beam resists. One layer of poly (methyl methacrylate) (495K PMMA with 2% Anisole, ~100nm) and on top, another layer of methyl methacrylate (950K MMA with 2% Anisole, ~70nm). On top of the resists, a very thin layer (~5-10nm) of Chromium (Cr) was deposited as a charge dissipation layer. Electron Beam Lithography (EBL, JEOL JBX-6000FS) was used to pattern the double layer resist. After patterning, the sample was developed at a cold temperature (4ºC) in a solution with 1-part Methyl Isobutyl Ketone (MIBK) and 3-part Isopropyl alcohol (IPA) to remove the exposed resists. Then a thin layer (~5nm) of Cr and 35nm thick Gold (Au) layer was deposited using thermal evaporation (Edwards Auto 306). The lift-off was done by soaking the sample in hot acetone for 12 hours, followed by an acetone and IPA rinse.

**Graphene Transfer**

We used a CVD graphene sample on a Cu foil (covered both sides). We cut a small piece, large enough to cover the whole area of the fabricated device and put it on a glass slide. We used one drop of water on the glass slide to hold the small piece in its place. Then we spin-coated 950K A4 PMMA layer (at 2500 RPM for 1 minute) on top of the graphene piece to protect the graphene layer on that side from the following processing steps. Then we flip the small piece and expose the bare graphene side of the Cu foil. We take another glass slide and use Kapton tape on all four sides of the piece to hold it on the glass slide. Then put it inside oxygen ($O_2$) plasma etcher for 15 minutes to etch the graphene off that side of the Cu foil. After the process is done, we remove the graphene-covered Cu foil (on one side now) and cut off the four edges of the foil as they were protected by the Kapton tape during the plasma etching and still have graphene on both sides. Then we put the graphene-covered Cu foil (Copper, Cu, side down) in a copper etchant (a mixture of $CuCl_2$ and HCl). After the etching is completed, we can see a translucent piece of graphene (covered with PMMA on top) floating on the etchant. We scooped the floating graphene piece from the etchant with a small piece of $SiO_2$/Si wafer and put it inside a Deionized (DI) water bath. We provide three separate DI water baths (each for 5 minutes) to the graphene piece to remove all the Cu etchants. After the last bath, we move the graphene sample to another DI water bath. Then we take our device (where we want to transfer graphene) and scoop the graphene sample out of the DI water with it (transferring graphene on our device). Then we gently blow the sample with an $N_2$ gun (optional as it may tear graphene) to remove any water puddle underneath the graphene and then let it dry on its own for 24 hours. After that, we put the sample in an acetone bath for 5 minutes and then rinsed it with acetone and IPA to remove all the PMMA from the top. Then blow-dry with an $N_2$ gun. Raman spectrum of the monolayer graphene is presented in Figure S16.

**Device reflection spectra measurement**

Bruker Vertex 70 FTIR spectrometer connected to a Hyperion 2000 microscope is used to measure the reflection spectra of the fabricated devices. A linear polarizer was used on the optical light path to align the light polarization along the long axis of the nanobar. The reflected light from the sample was collected with a 15x objective (NA=0.4). A liquid Nitrogen cooled Mercury Cadmium Telluride (MCT) photodetector was used to detect the collected reflected light from the sample. A background spectrum of the sample substrate ($Al_2O_3$/Al/Si) is taken and subtracted from all the measurements to remove any effect of the substrate.

**Degenerate Pump-Probe Measurement**

The ultrafast self-modulation is demonstrated by a degenerate (same color) pump-probe measurement. The measurement setup is illustrated in Figure 4a. A femtosecond Ytterbium fiber laser (Menlo Systems GmbH) is used to generate the pump and probe pulses at 1.035μm with a 100MHz repetition rate and ~110fs pulse width. The pump beam is focused on the sample at a 45° angle with a parabolic mirror, and the probe beam is focused at a normal angle on the sample with a ZnSe objective lens (NA=0.13). The reflected probe pulse is shrunk and focused with two plano-convex lenses on a biased free-space InGaAs photodetector (DET08CL, 5GHz, Thorlabs Inc.) and displayed by a mixed domain oscilloscope (DSA-X 91604A, Agilent Technologies) with 16GHz bandwidth. As the bandwidth and the response time of the photodetector is not enough to resolve the changes in the reflection of the probe pulses, we used a chopper (chopping frequency ~1200Hz) on the pump path and used it as a reference (5f option selected in chopper controller) for a phase-sensitive lock-in-amplifier. The photodetector output is connected to the lock-in-amplifier, and the DC output of the lock-in-amplifier is connected with a source meter and ultimately, the source meter reading is stored automatically by a computer using the MATLAB instrument control toolbox. An optical delay line controls the probe pulse delay (ODL 220, Thorlabs Inc.) with respect to the pump pulse. The movement of the ODL and the data recording from the lock-in-amplifier is synchronized and automated with the computer so that first the ODL is moved to change the delay and then a data is recorded.

## Acknowledgments

The research was supported in part by the AFOSR DURIP under grant FA9550-16-1-0183 and NSF 2DCC under grant DMR-1539116. Access to the NanoFab and/or EMC was supported, in part, by NSF contract ECCS-1542160.

## Conflict of Interest

The authors declare no conflicts of interest.

## Contributions

Y. Y. conceived the idea and supervised the study. M.Z.R. performed device design and characterization. M.Z.R., J.B., and J.J. fabricated the device. M.Z.R. and A.B. built the saturable absorption and pump-probe measurement setup. M.Z.R carried out the measurements and data analysis. M.Z.R. and Y.Y. wrote the manuscript, and all authors contributed to the manuscript.

# Supplementary Materials

Table of Contents

## S1. Plasmonic Metasurface and GPMSA

The multiple reflections in the Fabri-Perot cavity can be modeled with the modified Fresnel equations presented below[1],

$$r_{12} = \frac{n_1 - n_2 - \sigma_M Z_o}{n_1 + n_2 + \sigma_M Z_o} \tag{4}$$

$$r_{21} = \frac{n_2 - n_1 - \sigma_M Z_o}{n_2 + n_1 + \sigma_M Z_o} \tag{5}$$

$$t_{21} = \frac{2n_1}{n_1 + n_2 + \sigma_M Z_o} \tag{6}$$

$$t_{21} = \frac{2n_2}{n_1 + n_2 + \sigma_M Z_o} \tag{7}$$

$$r_A = \frac{E_R}{E_i} = \frac{r_{12} + (r_{12}r_{21} - r_{12}r_{21})r_{23}e^{i2kd}}{1 - r_{21}r_{23}e^{i2kd}} \tag{8}$$

$$r_{12} + (1 + r_{12} + r_{21})r_{23}e^{i2kd} = 0 \tag{9}$$

Equation (4)) - (7)) are the modified Fresnel equations. Furthermore, equation (8) represents the overall reflection from the device. If we set equation (8)) to zero, which means no reflection and all the light is absorbed, we can obtain equation (9)). This equation is the perfect absorption condition.

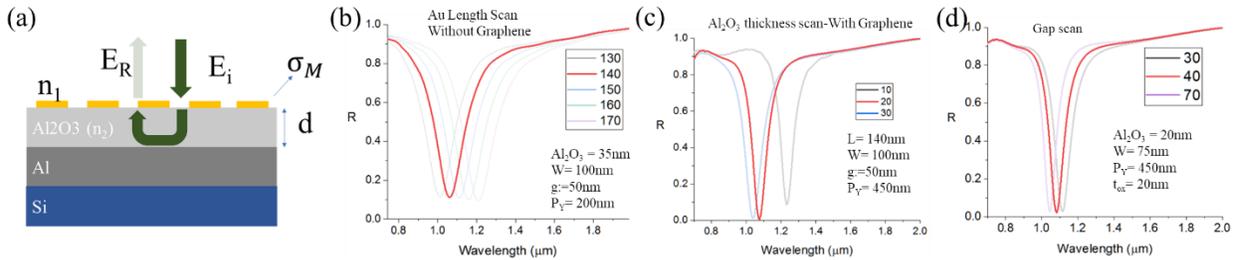

Figure S6: FDTD simulation to obtain the best graphene-plasmonic hybrid metasurface (GPMSA) design (a) Plasmonic metasurface design with Fabry-Perot cavity mode (b) Scanning the Au nanobar length to find the length corresponding to the resonance around 1.035μm without graphene (c) Al₂O₃ thickness scan to find which thickness provides the best absorption at 1.035μm. (d) Reflection spectra for different gaps between two nanobars.

The perfect absorption condition (eq. 6) depends on the metasurface conductivity ($\sigma_M$), dielectric layer thickness (d), complex dispersive refractive index ($n_2$) of dielectric layer, and wavelength of operation. We can modify the metasurface conductivity by changing the Au- nanobar length, width, thickness, gap between and period in y-axis ($P_Y$). Adding graphene on top is also going to change the metasurface conductivity depending on the quality of graphene, chemical doping, number of layers and so on. Furthermore, we can change the dielectric layer thickness to obtain perfect absorption in desired wavelength, which in this case is the wavelength of our laser in the lab at 1.035μm.

We performed FDTD simulation to optimize design parameters and achieve large absorption and high nearfield enhancement in the gap at resonance wavelengths. In Figure S6(b-d), this optimization process is highlighted with some selected plots. In Figure 1a, the near field enhancement inside the gap is presented. It shows a large enhancement of over 1000 at the laser wavelength. Also, at the laser wavelength, the absorption of the final device is around 99% (shown in Figure S6d). For near perfect absorption, the structure parameters range is as follows: length of Au nanobar, L = 130nm; width of Au nanobar, w = 75nm to 100nm; gap between nanogap, G=30 to 70nm; thickness of Au, $t_{AU}$= 40nm; $Al_2O_3$ dielectric layer thickness, $t_{ox}$= 20 to 30nm.

After the device fabrication, the structural parameters were changed and consequently the spectral response of the devices deviated from the ideal condition that we simulated here. For fabricated GPMSA devices, the best minimum reflection at the resonance (at around 1.035μm) we could achieve is around ~20% (As shown in the main manuscript Figure 3b). To recreate this in simulation we also changed the device parameters to achieve ~20% minimum reflection at the resonance (as shown in Figure 2c and all other simulations presented in the main manuscript accordingly).

So, GPMSA device parameters used in all simulations presented in this manuscript are Au nanobar length, L=120nm, width, W=100nm, gap, g=30nm, Au nanobar thickness, $t_{Au}$=40nm, x-period, $P_x$=150nm, y-period, $P_y$=450nm and $Al_2O_3$ thickness, $t_{AlO}$=40nm.

## S2. Time dependent optical conductivity of Graphene and FDTD simulation of GPMSA

From the coupled differential equations mentioned in the main manuscript, equation (1) and (2), we can calculate the time-dependent occupation probability in conduction band (CB), $f_c(t,\omega)$, and valence band (VB), $f_V(t,\omega)$. Then we can calculate the time-dependent non-equilibrium carrier concentration, $N(t,\omega)$, in the conduction band, within the energy band covered by the laser bandwidth, considering a certain relaxation time ($\tau_1$), with the following equation,

$$N(t,\omega) = \left(f_V(t,\omega) - f_c(t,\omega)\right) * D(E) \tag{S10}$$

The non-equilibrium photoexcited carrier concentration, $N(t,\omega)$, in the conduction band, at the laser wavelength (1.035μm), and subsequent relaxation due to carrier-carrier and carrier-optical phonon scattering is presented in Figure S7.

Graphene has a constant absorption of $\pi\alpha_f$ or 2.3% at all wavelengths. As the carriers accumulate in the conduction band from the valence band, the absorption decreases from 2.3%. We can account for the changing absorption, $A(t,\omega)$, by considering the time-dependent occupation probability of electrons in CB ($f_C(t,\omega)$) and VB ($f_V(t,\omega)$),

$$A(t,\omega) = \pi\alpha_f \left(f_V(t,\omega) - f_c(t,\omega)\right) \tag{S11}$$

Here, $A(t,\omega)$ is the time-dependent absorption of graphene. We can convert this absorption into an absorption coefficient, $\alpha(t,\omega)$, by considering a thickness (d) of 0.3nm for the graphene monolayer. Please see discussion in supplementary section S4 for absorption coefficient calculation of monolayer graphene. For these simulations, we take graphene absorption coefficient long before pumping, $\alpha(-\infty,\omega)$, as the initial or base value and then we try to calculate the changes in the absorption coefficient due to laser pumping, $\Delta\alpha(t,\omega) = \alpha(t,\omega) - \alpha(-\infty,\omega)$. From the changes in time-dependent absorption coefficient, $\Delta\alpha(t,\omega)$, the change in complex refractive index, $\Delta n_T(t,\omega) = \Delta n(t,\omega) + i\Delta k(t,\omega)$, of graphene can be calculated with the help of the Kramers-kroning model[2],

$$\Delta k(t,\omega) = c\frac{\Delta\alpha(t,\omega)}{2\omega} \tag{S12}$$

$$\Delta n(t,\omega) = \frac{c}{\pi}\,p.v\int_{\Omega_l}^{\Omega_m}\frac{\Delta\alpha(t,\Omega)}{\Omega^2 - \omega^2}\,d\Omega$$

$$\Delta n(t,\omega) = \frac{c}{\pi}\left[\int_{\Omega_l}^{\omega-}\frac{\Delta\alpha(t,\Omega)}{\Omega^2 - \omega^2}\,d\Omega + \int_{\omega_+}^{\Omega_l}\frac{\Delta\alpha(t,\Omega)}{\Omega^2 - \omega^2}\,d\Omega\right] \tag{S13}$$

Here, KK model relates the change in imaginary part ($\Delta k(t,\omega)$) and the real part ($\Delta n(t,\omega)$) of graphene's complex refractive index with its time dependent change in absorption coefficient, $\Delta\alpha(t,\omega)$. To calculate the effective total complex refractive index upon laser excitation, we extract the optical surface conductivity, $\sigma_S(-\infty,\omega)$, of non-pumped graphene from FDTD numerical model in Lumerical Inc. FDTD simulation software. From $\sigma_S(-\infty,\omega)$, we can calculate in plane ($\parallel$) complex permittivity, $\varepsilon(-\infty,\omega) = 1 + \frac{i\sigma_S(-\infty,\omega)}{\varepsilon_0\omega d}$, and then complex refractive index, $n_T(-\infty,\omega) = \sqrt{\varepsilon(-\infty,\omega)}$, of non-pumped graphene. Then we add the time dependent change in complex refractive index, $\Delta n_T(t,\omega) = \Delta n(t,\omega) + i\Delta k(t,\omega)$, to obtain the total complex refractive index of graphene,

$$n_T(t,\omega) = n_T(-\infty,\omega) + \Delta n_T(t,\omega) \tag{S14}$$

$n_T(t,\omega)$ includes the instantaneous time dependent index change due to laser pumping. The complex permittivity, $\varepsilon(t,\omega)$, of the graphene is then calculated from the complex index, $\varepsilon(t,\omega) = n_T(t,\omega)^2 = (n(t,\omega) + ik(t,\omega))^2$. Time dependent optical surface conductivity, $\sigma_S(t,\omega)$, of the graphene can be calculated from the in plane ($\parallel$) complex permittivity of graphene as follows,

$$\varepsilon(t,\omega) = 1 + \frac{i\sigma_S(t,\omega)}{\varepsilon_0\omega d} \tag{S15}$$

Using this instantaneous change in graphene's optical surface conductivity, $\sigma_S(t,\omega)$, we can perform simulation of GPMSA using the 2D surface conductivity model in Lumerical FDTD and find the time-dependent reflection spectra, as shown in Figure 2d and Figure S8a, of our device.

We sample optical surface conductivity ($\sigma_S(t,\omega)$) of graphene at different points of time in accordance with the change of non-equilibrium carrier accumulation, $N(t,\omega)$, in graphene CB (which is responsible for the instantaneous change in optical conductivity). Then we use these $\sigma_S(t,\omega)$, at different time, as conductivity input for the graphene 2D conductivity model in Lumerical FDTD to do the time dependent FDTD simulation of our device. The process of selecting optical surface conductivity at different points of time is depicted in Figure S7. Pump fluence used in this simulation is $0.0566\mu J/cm^2$.

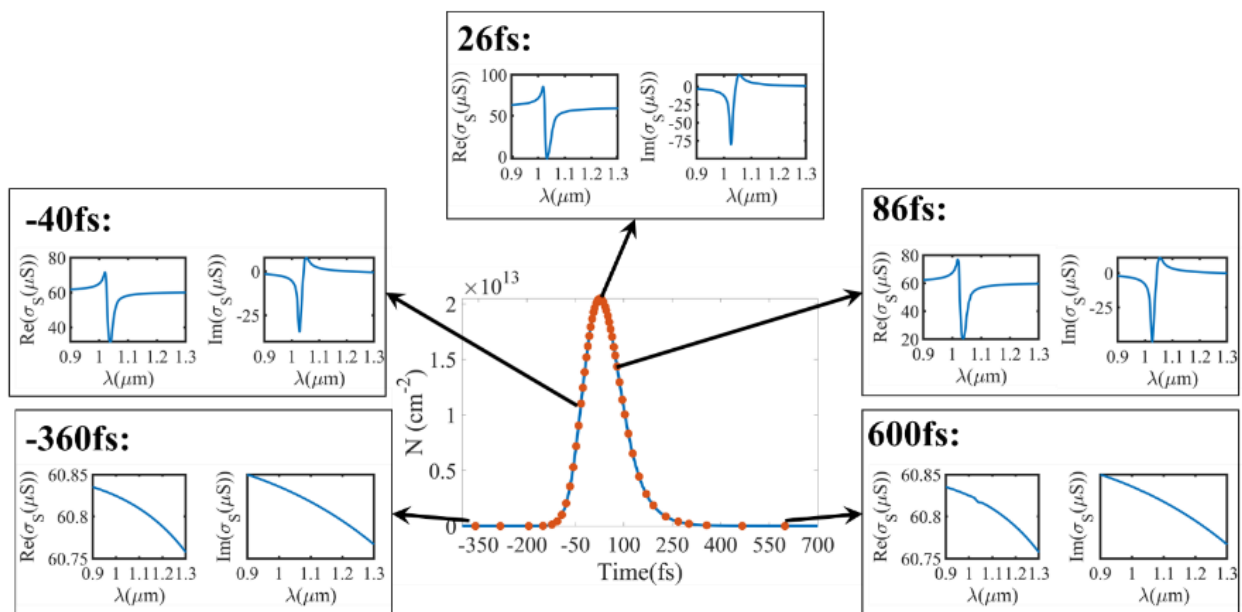

Figure S7: Process of selecting different point of time on the carrier density plot and finding graphene conductivity at each of those times. FDTD simulation is done with corresponding surface conductivity of graphene for each of the selected points. This way we can perform a time-dependent simulation for our graphene-plasmonic hybrid metasurface device (GPMSA).

Simulation results depicting the time evolution of reflection spectra, R, and reflection modulation ($\Delta R$) with changing optical surface conductivity of graphene in GPMSA is presented in Figure S8b. This simulation is done considering a plasmonic metasurface with gap size 30nm and incident pump fluence of $56.6 \text{nJ/cm}^2$.

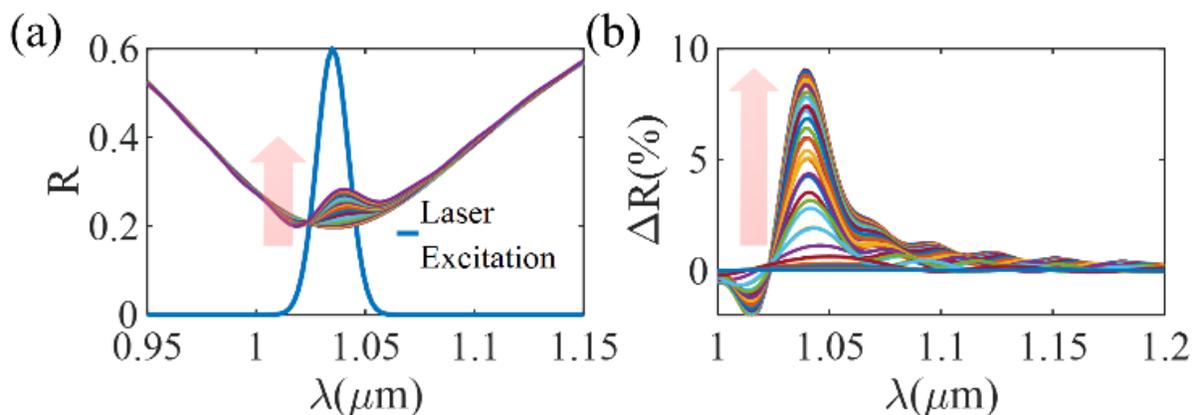

Figure S8: (a) Change in the hybrid metasurface reflection spectra with time (within 10-20fs) after laser excitation. The red arrow indicates the change direction with increasing time. (b) Gradual evolution of reflection modulation in GPMSA with time (within 10-20fs) after laser excitation.

## S3. Time dependent Fullwave FDTD Simulation in a Self-Consistent way

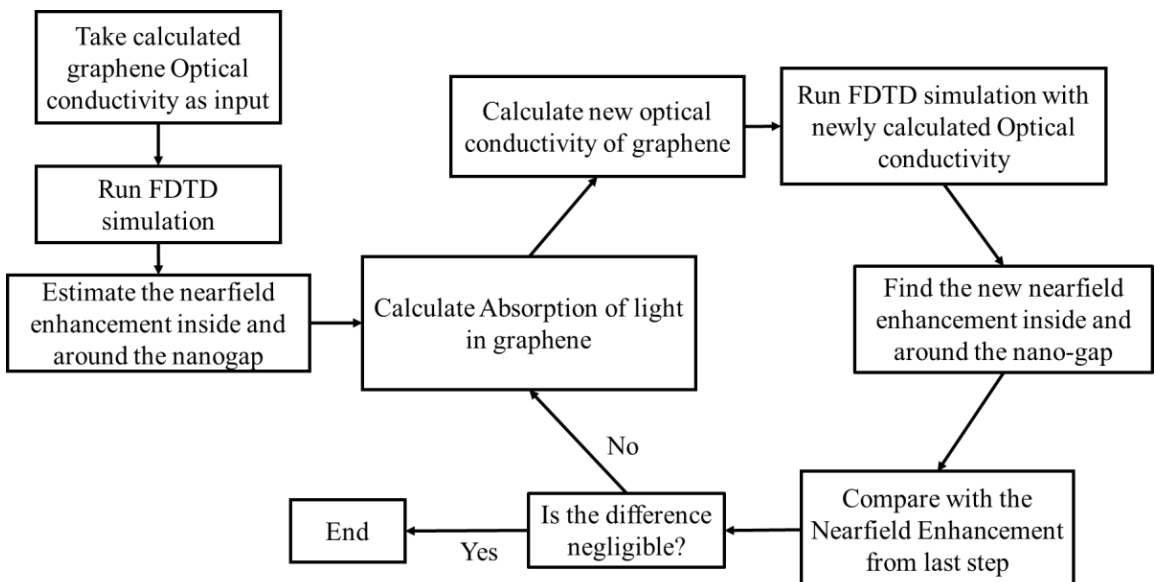

Figure S9: Flow chart of self-consistent FDTD simulation

The process of time dependent Fullwave FDTD simulation in a self-consistent way is shown in Figure S9. The process is carried out at each of the preselected time points of the incident pulse (Figure S7). Carrier density at the half of the incident photon energy and corresponding graphene optical conductivity is calculated based on the process described in section S2 which we feed it back to FDTD simulation to obtain the near-field enhancement inside the nanogaps and the reflection spectra of GPMSA. This iterative process continues until the simulated nearfield enhancement inside the nanogaps match with the previously simulated near-field enhancement.

Monolayer graphene on the plasmonic metasurface is modelled considering four different regions according to the enhanced near field in different regions of the hybrid metasurface. These regions are indicated by R1, R2, R3 and R4. The graphene around the nano-gap (R1, R2 and R3) experiences much more near-field enhancements compared to the surrounding graphene (R4) on the other parts of the metasurface. For a plasmonic nano-bar design with gap 30nm and width of 100nm; area of R1 is 15nmx 105nm, R2 is 7.5nmx 105nm and R3 is 120nmx 31nm and the rest is R4. These are illustrated in Figure S10.

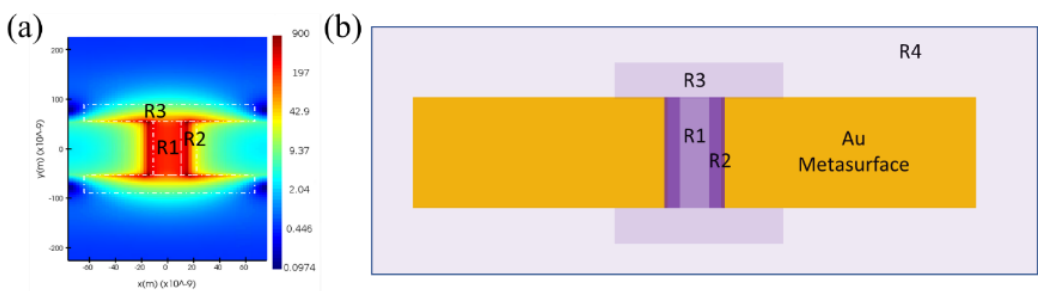

Figure S10: (a) Different regions of graphene experience different near field enhancement. (b) Based on (a) graphene is modeled in four different regions. R1- Middle of nano gap, R2- Near the edge of nano-gap, R3- On the side of nano-gap and edge of nanobar, R4-rest of the graphene regions covering the GPMSA and surroundings.

The average near field enhancement is calculated for the regions discussed in Figure S10. This near field enhancement is multiplied with the incident pulse intensity to get the new incident peak intensity. This new pulse intensity is used to calculate graphene's new optical conductivity for each indicated regions (R1, R2 and R3). Then we use this new optical conductivity to simulate our device in FDTD again to obtain the new near field enhancement in those regions. Comparing the difference in the near field enhancement, between this step and the previous, we can set a stop to this loop. The entire process is illustrated in Figure S11 for incident pump fluence of 0.0566μJ/cm2.

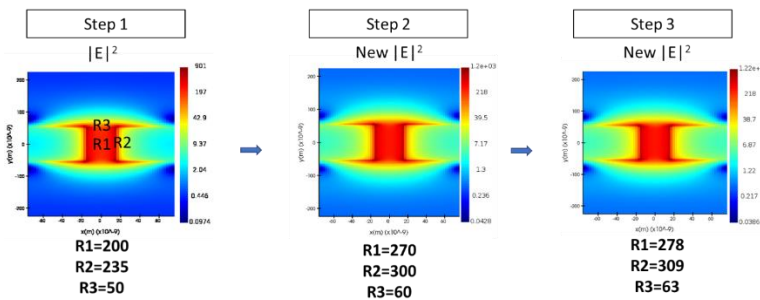

Figure S11: Step 1: Near field enhancement around the nanogap for the input optical conductivity of graphene. The average near field enhancement for different regions are indicated below the near field-enhancement for each step. New optical conductivity is calculated based on the near-field enhancement obtained in step 1 . Step 2- the resulted near field enhancement after using the newly calculated optical conductivity of graphene is shown here. As the near field enhancement in different regions are quite different compared to step 1, we use this new near field enhancement to calculate the new optical conductivity of graphene. Step 3- We use the newly calculated optical conductivity to do another simulation of our device. It seems the near field enhancement after this simulation is very close to the results in step 2.

### S4. Graphene n and k calculation

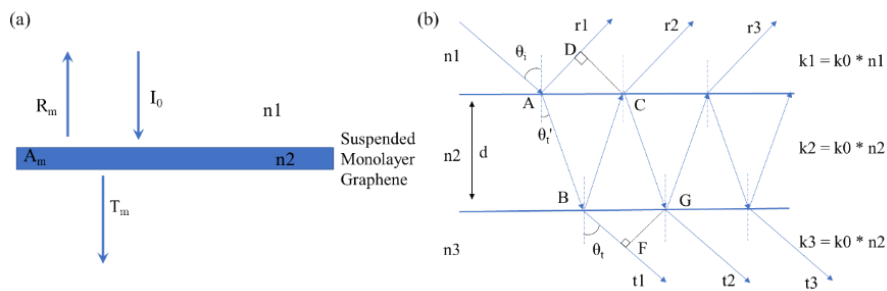

Figure S12: (a) Suspended monolayer graphene surrounded by air. (b) Reflection, transmission, and absorption inside monolayer graphene depends on the multiple reflection in the cavity formed inside graphene.

The transmission, reflection and absorption from the monolayer graphene depends on the multiple reflections happening inside the cavity formed inside graphene surrounded by air, illustrated in Figure S12a and b.

Total reflection coefficient considering multiple reflection inside the cavity,

$$r_{tot} = r_{12} + t_{12}r_{23}t_{21}e^{i(2\varphi_h - \triangle\varphi_r)}\frac{1}{1 - r_{21}r_{23}e^{i(2\varphi_h - \triangle\varphi_r)}}$$

Total transmission coefficient considering multiple reflection inside the cavity,

$$t_{tot} = t_{12}\, e^{i\varphi_h}\, t_{23} \frac{1}{1 - r_{23}r_{21}e^{i(2\varphi_h - \triangle\varphi_t)}}$$

Here,

$r_{ij}$ = reflection coefficient of an interface when light going from i medium to j medium

$t_{ij}$ = transmission coefficient of an interface when light going from i medium to j medium

Path traversed inside material with index n2 (graphene) is AB and BC. Phase accumulated by AB and BC is:

$$\varphi_h = k_2 * d/ \cos(\theta_t')$$

Phase difference introduced due to path difference is provided as follows

$$\Delta\varphi_r = k_1 *2d \tan(\theta_t') \sin(\theta_i) = 2k_2d\ (\sin^2(\theta_t')/\cos(\theta_t'))$$
$$\Delta\varphi_t = k_1 *2d \tan(\theta_t') \sin(\theta_i) = 2k_2d\ (\sin^2(\theta_t')/\cos(\theta_t'))$$

For vertical incidence,

$$\varphi_h = k_2 * d \text{ and } \Delta\varphi_r = \Delta\varphi_t = 0$$

So, total reflection and transmission coefficient becomes

$$r_{tot} = r_{12} + t_{12}r_{21}t_{21}e^{i(2k_2d)} \frac{1}{1 - r_{21}r_{21}e^{i(2k_2d)}} \tag{S16}$$

$$t_{tot} = t_{12}\, e^{ik_2d}\, t_{21} \frac{1}{1 - r_{21}r_{21}e^{i(2k_2d)}} \tag{S17}$$

Here, medium 3 and 1 are same, air. So, n1 =1 and n2= n+ik.

Reflection and transmission coefficient from ach interface is provided below

$$r_{12} = \frac{n_1 - n_2}{n_1 + n_2} = \frac{1 - (n+ik)}{1 + (n+ik)}$$
$$r_{21} = \frac{n_2 - n_1}{n_1 + n_2} = \frac{(n+ik) - 1}{1 + (n+ik)}$$
$$t_{12} = \frac{2n_1}{n_1 + n_2} = \frac{2}{1 + (n+ik)}$$
$$t_{21} = \frac{2n_2}{n_1 + n_2} = \frac{2(n+ik)}{1 + (n+ik)}$$

And propagation constant or wavenumber inside graphene is

$$k_2 = n_2 * k_0 = (n+ik)\frac{2\pi}{\lambda}$$

So, $e^{i(2k_2d)} = e^{i(2(n+ik)\frac{2\pi}{\lambda}d)} = e^{2(in-k)\frac{2\pi}{\lambda}d} = e^{i2n\frac{2\pi}{\lambda}d - 2k\frac{2\pi}{\lambda}d} = e^{i2n\frac{2\pi}{\lambda}d - \alpha d} = e^{i2n\frac{2\pi}{\lambda}d}e^{-\alpha d}$

Absorption coefficient of graphene,

$$\alpha = 2k\frac{2\pi}{\lambda} = 2k\frac{2\pi f}{c} = 2k\frac{\omega}{c}$$

Graphene material properties, under laser excitation, is derived from the coupled differential equation in main manuscript.

## S5. Optical conductivity of graphene at different pump fluences

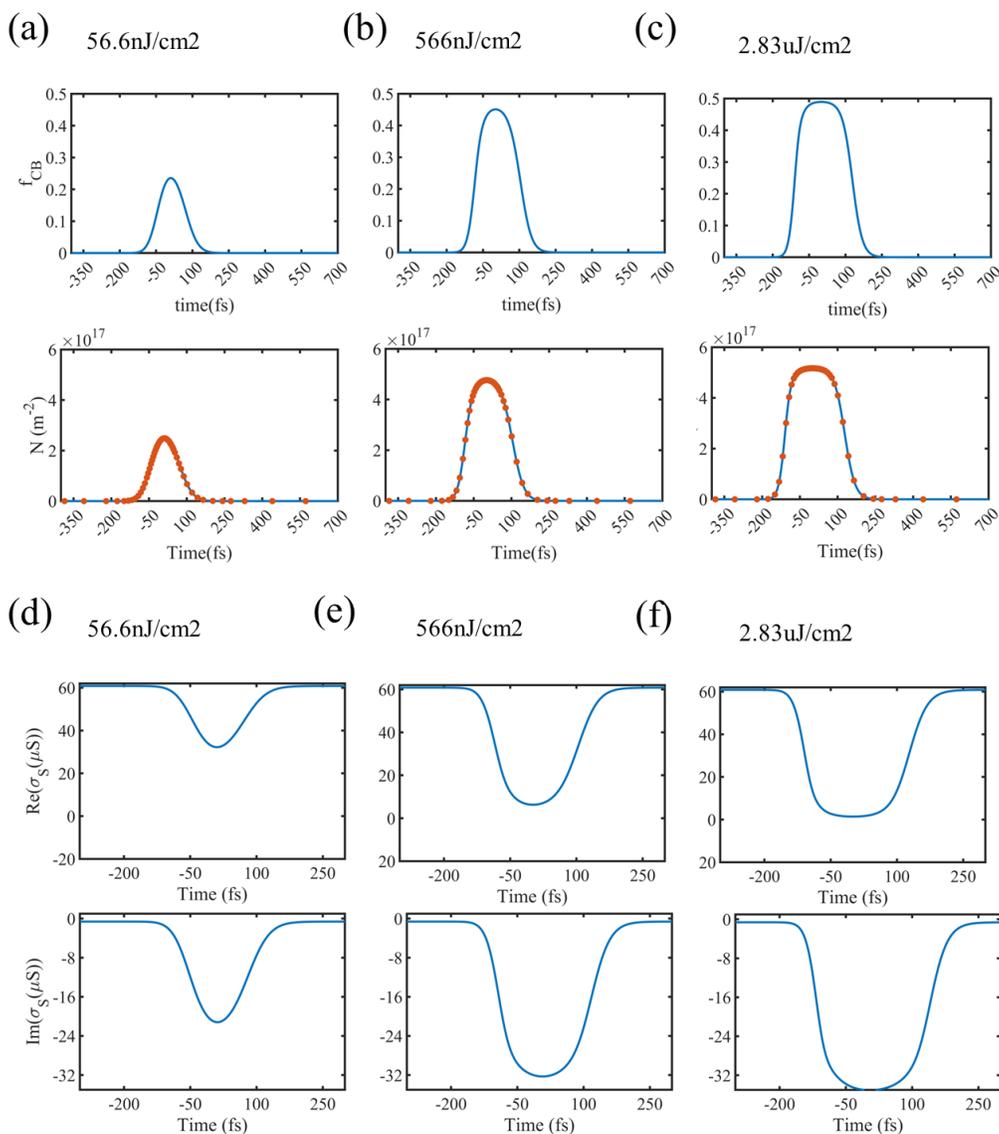

Figure S13: After laser excitation, the graphene experiences changes in its optical properties. The changes occur at the energy band corresponding to the laser wavelength bandwidth (1.035μm ± 17nm). (a-c) Presents changes in occupation probability of photoexcited carriers in graphene CB ($f_{CB}$) and non-equilibrium carrier density ($N(m^{-2})$), at different pump fluences, within the energy band in CB corresponding to the incident laser's wavelength bandwidth. (d-f) Presents changes in real and imaginary part of graphene's optical conductivity ($\sigma_S$), at different pump fluences, within the energy band in CB corresponding to the incident laser's wavelength bandwidth.

## S6. Calculation of Absorption and Absorption Enhancement in Graphene

Absorption is calculated utilizing FDTD simulation software from Lumerical Inc. We setup an FDTD simulation and to calculate the absorption within a certain region of graphene we follow the below outlined procedure.

1. Setup FDTD simulation with advanced absorption analysis group.
2. Enclose the region of interest and a small part of surrounding region (at least one cell larger) with this advanced absorption analysis group. This analysis group contains an index monitor and electric field monitor.
3. Run the simulation. It will provide the |E|$^2$ at different position of the simulation region inside the enclosed analysis group.
4. Using the formula below the analysis group calculates the power absorbed per unit volume, Pabs, at different position of the enclosed region. Pabs.x, Pabs.y and Pabs.z. It depends on the imaginary part of permittivity and the electric-field intensity,

$$P_{abs} = -0.5 \, \varepsilon_0 \, \omega |E|^2 \, Im(\varepsilon)$$

5. Create a filter (F) for the specific material we want to calculate the absorption for. We can use the material parameter used in the simulation (obtained from the index monitor) to find the exact region enclosed by the material. This is in terms of x,y, and z.
6. Then integrate (Pabs x Pabs x F) over 3D (x,y and z) space with corresponding position vector defined in Pabs.x, Pabs.y and Pabs.z to get the absorption,

$$\text{Absorption} = \iiint\limits_{x_i y_j z_k}^{x_{N_1} y_{N_2} z_{N_3}} P_{abs} P_{abs} \, F \, dx \, dy \, dz$$

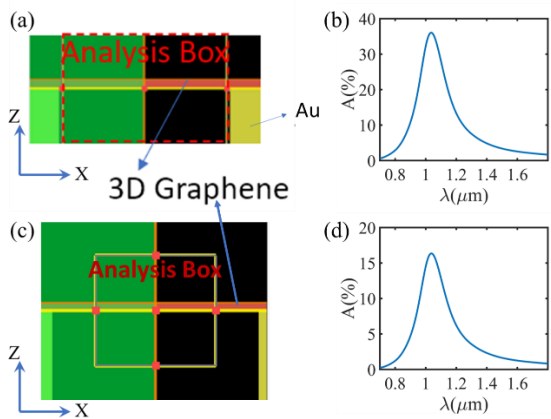

Figure S14: (a) Enclosed region inside the analysis box to calculate the absorption in graphene within the nano-gap (29.4nm x 107nm). (b) Calculated absorption of light inside graphene within the whole nano-gap region. (b) Enclosed region covering the middle of the nano-gap (17nm x 107nm) (d) Absorption of light in graphene within the region enclosed in (c).

Absorption enhancement is obtained by comparing the absorption of a certain region with the absorption of light in a suspended graphene (2.3%) having the same area. So, the absorption enhancement is obtained by,

Absorption Enhancement= A (%) / 2.3% / (A$_C$/A$_T$)

Here, A (%) is the absorption within the covered area ($A_C$), $A_T$ is the total simulation area, in our case its 160nm x 450nm.

So, maximum absorption enhancement within the middle of the nano-gap region (Region 2: 17nm/107nm) (Figure S14c) is around 280 times. And the maximum average absorption enhancement in the rest of the gap region (Region 1: 13nmx107nm) is around 450 times. Illustrated in Figure S15.

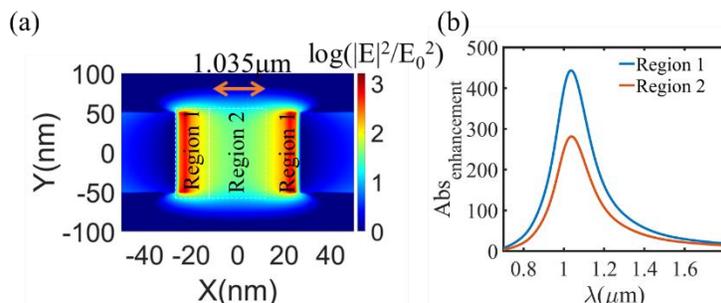

Figure S15: (a) Different region inside the nano-gap showing different near-field intensities (b) Absorption enhancement inside different region inside the nano-gap.

## S7. Simulation of Reference Device- Graphene on Substrate

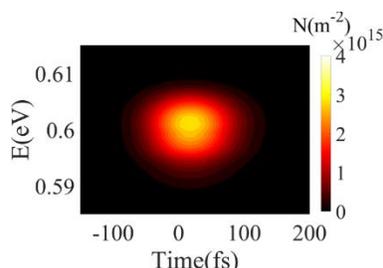

Figure S16: Simulated 2D contour plot of the number of excited non-equilibrium carriers, $N(m^{-2})$, within the narrow energy band (around $\frac{1}{2}\hbar\omega$) of the conduction band of graphene at 112nJ/cm2 pump fluence in suspended graphene. The X-axis represents time, and Y-axis represents the relative energy of the graphene conduction band calculated from the edge

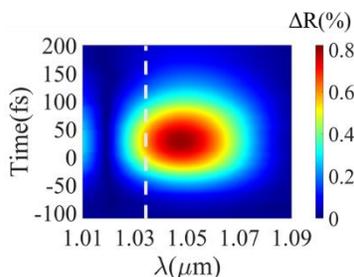

Figure S17: Simulated reflection modulation ($\Delta R(\%)$) from graphene lying on $Al_2O_3$/Al/Si substrate for 5.66μJ/cm² pump fluence.

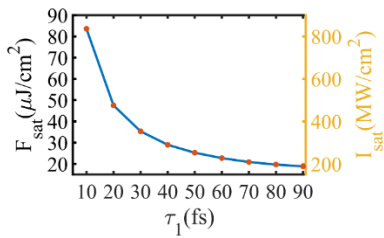

Figure S18: Simulated saturation fluence ($F_{sat}$) vs relaxation time constants ($\tau_1$) for reference device, graphene on substrate ($Al_2O_3$/Al/Si).

## S8. Device Simulation with different carrier relaxation times

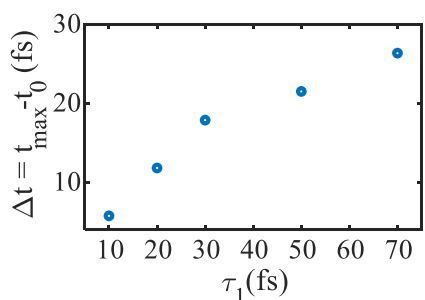

Figure S19: Carrier relaxation time in graphene, $\tau_1$, vs. the time it takes for the reflection modulation to reach maximum after laser pulse maxima. Here, pulse width of laser is 110fs and pump fluence is 56.6nJ/cm².

It is evident, from Figure 2g, that the reflection modulation reaches its maxima at a later time compared to the incident pulse maxima at time zero. The time delay between the maximum reflection modulation and the peak of incident pulse is because it takes some time for the non-equilibrium carriers to build up in the excitation energy band ($\frac{1}{2}$ of the photon energy, $\hbar\omega$ or ~0.6eV). The carrier build-up time is required because as the excited carriers occupy the empty states close to the excitation energy band, the e-e scattering[3] scatter them off from the occupied states; however, the scattering rate is slower compared to the carrier excitation rate, so eventually, the carriers build up. Consequently, the longer the ultrafast relaxation times of graphene ($\tau_1 < 50$fs in Figure 2g), the rising part of the reflection modulation becomes longer than the decaying part.

## S9. GPMSA Reflection modulation at off-resonance laser excitation

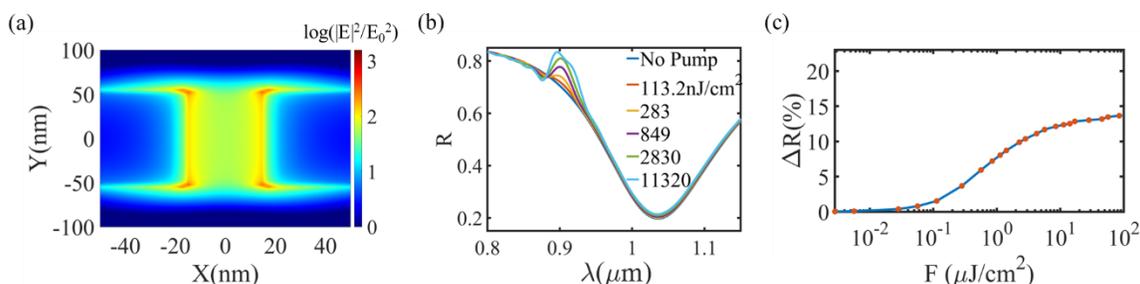

Figure S20: (a) Simulated Near-field enhancement inside nano gap of GPMSA at 900nm pumping (away from the resonance). (b) Reflection spectra modulation and (c) Reflection modulation, $\Delta R$ (%), of GPMSA device for different incident pumping fluences at 900nm (off-resonance pumping).

## S10. Graphene transfer and Raman Spectroscopy

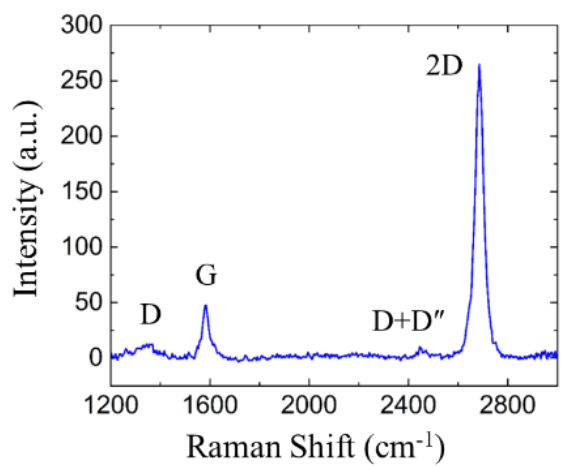

Figure S21: Raman Spectrum of the transferred graphene on the metasurface. The ratio between 2D and G peak is more than 2 which indicates monolayer graphene, and a very low D peak indicates high quality and defect-free graphene.

After graphene transfer, we verified the monolayer graphene by doing Raman spectroscopy, illustrated in Figure S21. The 2D peak, around $2700cm^{-1}$, is more than two times compared to the G peak, around $1580cm^{-1}$. Furthermore, the D peak is very small, indicating minimal defect or disorder in graphene ensuring high quality.

## S11. FTIR Setup and Reflection spectra

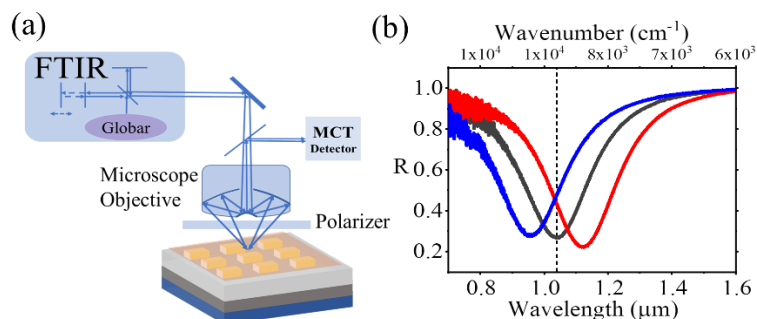

Figure S22: (a) FTIR, microscope and MCT detector setup for reflection spectra measurement. (b) Reflection spectra of different GPMSA devices with different resonance (working wavelength) wavelength.

## S12. Pump fluence dependent Reflection measurement setup

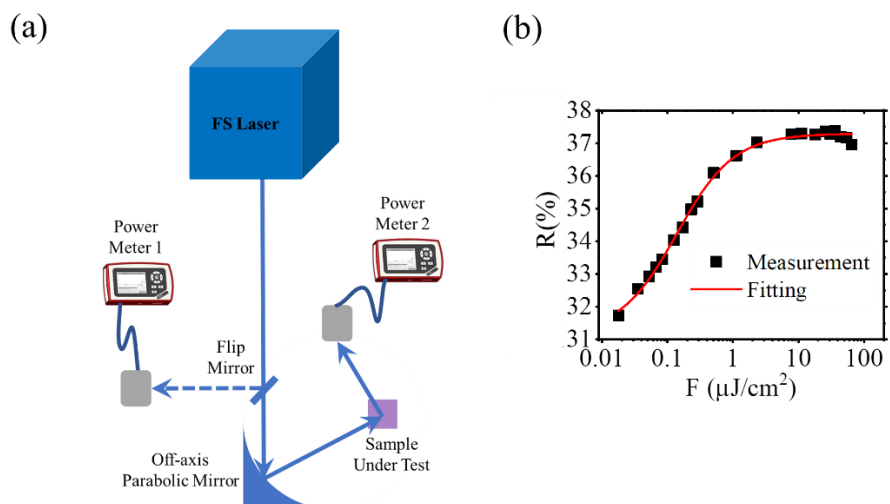

Figure S23: (a) Measurement setup to characterize the saturable absorption property of the device. A flip mirror in the incident light path is used to re-direct the beam toward the power meter 1 to monitor the intensity fluctuation in the laser. An off-axis parabolic mirror is used to focus the incident pump laser on the sample a second power meter is used to capture the reflected laser light. (b) Measured reflection, R (%) at 1.035μm, of GPMSA (same device as in Figure 3a) at different incident pump fluence and its fitted line.

The optical setup for the saturable absorption measurement is shown in Figure S23a. The setup comprises a femtosecond laser (Orang-Menlo System) with 100fs pulse width and 100MHz repetition rate at center wavelength 1.035μm and spectral broadening of ~17nm. The laser's output power can be changed from 0 to 1200mW by varying pump diode currents. The diameter of the focus point on the sample is around 150μm (supplementary S12). The fabricated GPMSA device area is 200μm x 200μm, so the laser beam is focused on the device's midpoint to capture almost all of the incoming light intensity. After the incident light is reflected off the sample, it is captured by a power meter (S302C- S302C-Thermal Power Sensor Head, Thorlabs Inc., power meter 2 in Figure S23a). First, we record the reflected power from the substrate and then we move to record the reflected power from the GPMSA. These recorded powers are used to calculate the reflection of the sample (GPMSA and reference device) at 1.035µm.

Figure S23b shows the measured reflection profile of Device 1 with respect to different incident pump fluence. This measured data is fitted with equation (3), and the obtained saturation fluence, $F_s$, of the device is ~100nJ/cm$^2$. The fitted curve is shown with solid red line.

## S13. Beam diameter calculations and measurements

The pump beam is focused using a 45° parabolic mirror with reflected focal length 8". The beam diameter at the focus point is calculated considering gaussian pulse shape. The wavelength of the pump light is 1.035µm and incident beam diameter is 2mm. Considering the gaussian laser beam, the diameter is calculated to be 135µm. But because the beam is falling on the device at 45°, the focus point becomes elliptical and one of the axes become longer and the other axis is around 135µm.

We confirmed the beam diameter at the focus point by focusing the beam on reflective Au circles (fabricated on glass substrate) of different diameters 50µm, 100µm, 150µm, 200µm and 250µm (these circles are used to measure beam diameters of other lasers also). For us, we focused the laser beam on the middle of the reflective circle of 200µm and 250µm diameters. Then collected the reflected beam in a power meter. Then scanned the beam on x-and y-axis of the circle with 10µm steps. We start counting the steps just before we see the decrease in signal in the power meter and continue counting the number of steps until we completely lost the signal on the power meter. It takes about 14 to 16 steps in this case. So, we approximate the shape of the focus point of beam to be circular with diameter around 150µm.

## S14. Reflection spectra of different devices before and after graphene transfer

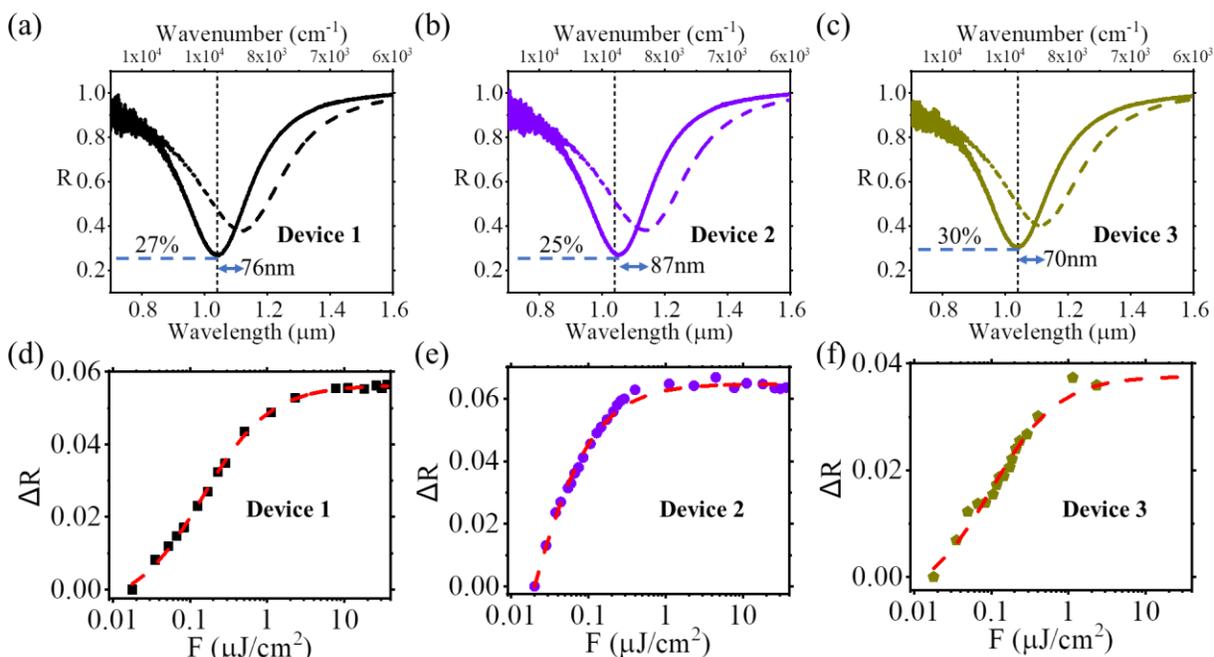

Figure S24: (a-c) FTIR spectra of different devices before and after graphene transfer. The different blue shifts of resonance, absorption and reflection modulation before and after graphene transfer is presented

in Table 1. (d-e) Measured reflection modulation profile of device 1 to 3. Structure parameters for Device 1: L=150nm, W=100nm, g=46nm, and $P_Y$=470nm (same device as in Figure 3c); Device 2: L=155nm, W=96nm, g=46nm, and $P_Y$=470nm and Device 3: L=150nm, W=95nm, g=46nm, and $P_Y$=470nm. For all GPMSA devices, $t_{Au}$=40nm, and $t_{AlO}$=20nm.

Table 1: Changes of different device response after graphene transfer

| Device | $\Delta R_{1.035\mu m} = R_{1.035\mu m}$ (w/o Gra.) - $R_{1.035\mu m}$ (w Gra.) | Absorption at resonance (A=1-R) | $\Delta\lambda = \lambda_{Resonance}$ (w/o Gra.) - $\lambda_{Resonance}$ (w Gra.) | Reflection Modulation, $\Delta R$(%) |
|---|---|---|---|---|
| 1 | 20% | 73% | 76nm | 5.5% |
| 2 | 22% | 75% | 87nm | 6.5% |
| 3 | 19% | 70% | 70nm | 3.7% |

The reflection spectra of all three GPMSA devices are presented in Figure S24(a-c). These three devices have slightly different structure parameters, but they all have resonance very close to 1.035μm, the operation wavelength of the laser (same as in Figure 3b). Furthermore, although these devices are all pumped around the resonance, the reflection modulation is different due to different blue shift of the device resonance after graphene transfer, different absorption at the resonance, and different reflection modulation at the incident laser wavelength before-after graphene transfer.

From Figure S24(a-c) and Table 1, we see that the device resonance blue shifts differently which results in different reflection modulation and absorption at pump wavelength (1.035μm). The maximum blue shift happens for device 2, then device 1 and then device 3. Consequently, the maximum reflection modulation is observed for Device 2, ~6.5% in Figure 3c, for device 1 it is around 5.5% and for device 3 its around 3.7%.

Although the net reflection modulation for GPMSA is larger than the reference device, they are quite small compared to other SA devices presented in literature. The small net reflection modulation, $\Delta R$ (%), can be explained with the similar logic presented in the context of Figure 2g, which states that the GPMSA device having an ultrafast recovery time will have small reflection modulation. The evidence for this ultrafast recovery is presented in the main text, where we discuss the transient behavior of GPMSA devices.

## S15. Pump-probe simulation

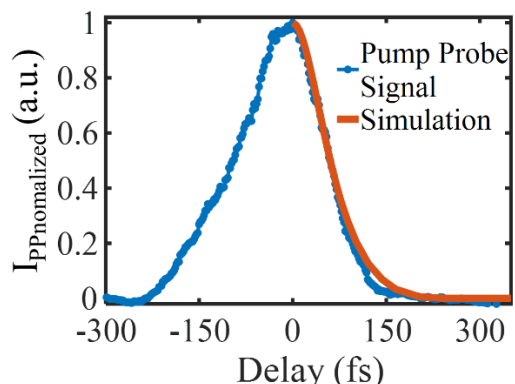

Figure S25: Measured pump-probe signal, normalized with max value, overlapped with the simulated pump-probe signal (Device 1) at 566nJ/cm² pump fluence.

## S16. Comparison of our work with reported works in literature

Here we compare the performance of GPMSA with different reported works in the literature. In Table S2, we have listed some of the saturable absorbers in literature and their performance parameters such as saturation fluence, recovery time, modulation depth, and insertion loss.

**Table S2: Comparison of our work with reported works in literature**

| | Wavelength (nm) | Saturation fluence | Saturation intensity | Recovery time | Modulation Depth | Insertion Loss |
|---|---|---|---|---|---|---|
| SESAM[4] | 1314 | 1.1 μJ/cm² | - | - | 4% | -10.2119dB |
| Monolayer Graphene[5] | 750 | 5.3 mJ/cm² | 5300 MW/m² | 100fs | 65.9% | 4.9592 dB |
| Multilayer Graphene[6] | 630 | 7.13 mJ/cm² (3 layer) | 7100 MW/m² (3 layers) | 0.21ps | ~65% (3 layer) | 4.34 dB |
| Graphene-Bi2Te3 heterostructure[7] | 1550 | 15-22 μJ/cm² | 72.9 to 109.5 GW/m² | 191-296fs | 11.5-41.88% (depending on coverage of $Bi_2Te_3$) | 2.5 to 5 dB |
| MoS₂ Nanoplatelets[8] | 1064 | 6.9 mJ/cm² | 8.7 MW/m² | - | 4.6% | 0.34 dB |
| WS₂ film[9] | 1560 | 17 μJ/cm² | 250 GW/m² | - | 1.20% | 0.18 dB |
| Few Layer PtSe₂[10] | 1064 | - | 0.346 GW/cm² | - | 26% | - |
| $Bi_2Te_3$ Topological Insulator film[11] | 1560 | 3.74 μJ/cm² | 26.7 MW/cm² | - | 5.7% | - |
| Large-Diameter SWCNT[12] | 2300 | 18 μJ/cm² | - | 220fs | 30% | - |
| Carbon-nanotubes[13] | 1500-1630 | 11.6 μJ/cm² | - | - | 9.182% | - |
| Ionic Liquid Gated CNT[14] | 1540 | - | - | 80 to 190fs (At different Applied voltages) | 0.7 to 3.2 % | - |
| Semiconductor Metasurface[15] | 830 | 96 μJ/cm² | 1.6 GW/cm² | ~6ps[16] | 25% | 1.5dB |
| Plasmonic-Metasurface[17] | 1555 | 0.64 J/cm² | - | - | 60% | 8.42 dB |
| **GPMSA- Our work** | **1035** | **~0.10 μJ/cm²** | **~1.0 MW/cm²** | **<60fs** | **20%** | **5dB** |

Modulation Depth = $(\Delta R/R)$ x100% and Insertion loss = $10 \log R$ dB. In Figure S26(a, b) we illustrate the comparison between different devices and GPMSA.

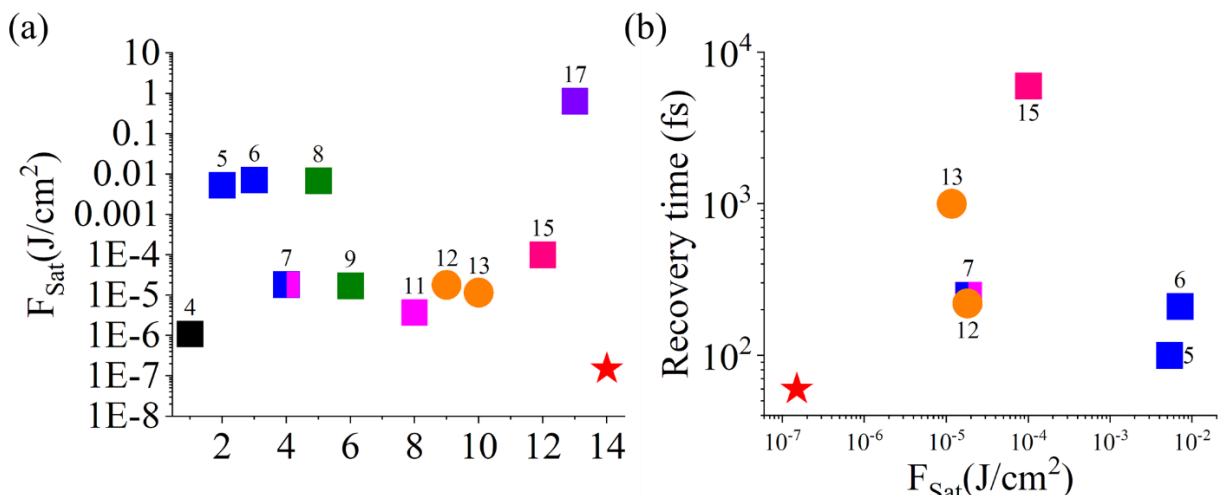

Figure S26: Our work in comparison to other reported works in literature. (a) Comparing Saturation fluence of different reports (b) Recovery time combined with saturation fluence. The red star represents our work. The reference number of the reported work is provided as the label.

## S17. Laser spectra measurement

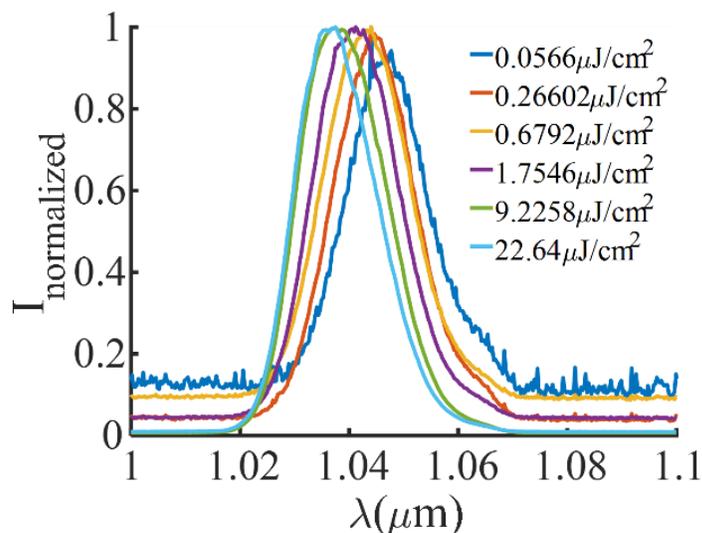

Figure S27: Laser spectra at different pumping fluence on the sample. Laser output power is changed by changing the amplifier pump current.

A femtosecond Ytterbium fiber laser (Menlo Systems GmbH) is used in our measurements. Around 10nm spectra blue shift is observed in the laser's spectra when output fluence is regulated by changing the laser's pump current (using internal control).

On-resonance devices (resonant around the laser wavelength) are pumped from ~0.04µJ/cm$^2$ to ~20µJ/cm$^2$, in this pump range the femtosecond laser spectra will blue shift by ~10nm. This blue shift causes ~0.1% reflection modulation (unrelated to saturable absorption) of the device, which is negligible considering the 6% reflection modulation obtained from the saturable absorption measurement. Furthermore, the saturation absorption for these devices happens ~$0.1 \mu J/cm^2$. In this pump range the femtosecond laser has <3nm blue shift. So, the measurement results obtained from the on-resonance devices are not affected by the wavelength shift of the laser.